\documentclass[preprint,12pt,numafflabel]{article}
\usepackage{amsmath, amssymb, amscd, amsthm, amsfonts}
\usepackage[title]{appendix}
\usepackage{graphicx}
\usepackage{hyperref}
\usepackage{color}
\usepackage{authblk}
\usepackage{subcaption}
\usepackage{float}
\usepackage[section]{placeins}
\usepackage{indentfirst}
\usepackage{comment}
\usepackage{diffcoeff}

\title{Temperature-Dependent Calibration Procedures for the Silicon Photomultiplier Readout of the Cosmic Ray Veto Detector for the Mu2e Experiment}
\author[1]{Lincoln Curtis}
\author[1]{E. Craig Dukes}
\author[1]{Ralf Ehrlich}
\author[1]{Josh Greaves}
\author[1]{Craig Group}
\author[1]{Karl Hardrick}
\author[1]{Tyler Horoho}
\author[2]{Yuri Oksuzian}
\author[3]{Paul Rubinov}
\author[1]{Matthew Solt}
\author[2]{Yongyi Wu}
\author[1]{Anran Zhao}
\affil[1]{University of Virginia, Charlottesville, VA 22904, USA}
\affil[2]{Argonne National Laboratory, Lemont, IL 60439, USA}
\affil[3]{Fermi National Accelerator Laboratory, Batavia, IL 60510, USA}
\date{\today}

\begin{document}
\maketitle

\begin{abstract}

The cosmic ray veto detector for the Mu2e experiment consists of scintillation bars embedded with wavelength-shifting fibers and read out by silicon photomultipliers (SiPMs).  In this manuscript the calibration procedures of the SiPMs are described including corrections for the temperature dependence of their light yield.  These corrections are needed as the SiPMs are not kept at a constant temperature due to the complexity and cost of implementing a cooling system on such a large detector.  
Rather, it was decided to monitor the temperature to allow the appropriate
corrections to be made.  The SiPM temperature dependence has been measured in a dedicated experiment and the calibration procedures were validated with data from production detectors awaiting installation at Fermilab.

\end{abstract}

\newpage
\pagenumbering{arabic}
\tableofcontents
\newpage

\section{Introduction}\label{sec:into}
The Mu2e experiment intends to search for the conversion of a negatively charged muon 
into an electron in the presence of an aluminum nucleus: 
$\mu^- {\rm Al} \rightarrow e^- {\rm Al}$~\cite{Mu2e:2014fns}.  In the Standard Model, 
this process is forbidden. In the extended Standard Model with non-zero 
neutrino masses, the ratio of conversion to capture is at an experimentally unobservable value of 
$R(\mu^-{\rm Al} \rightarrow e^-{\rm Al}) \simeq 10^{-52}$~\cite{Marciano:2008zz}.  Hence, any observation of this process would be unambiguous 
evidence of new physics.

For any sensitive search experiment, it is vital to keep backgrounds under control, and if possible to less than one event over the duration of the experiment.  Hence, the Mu2e apparatus has been 
designed to minimize backgrounds while keeping the signal efficiency high.  
An enormous simulation effort has been made to estimate the various backgrounds: 
see Ref.\,\cite{SU2020} for details.  The backgrounds come in three types: 
(1) stopped-muon-induced backgrounds; (2) beam-related backgrounds;
and (3) time-dependent backgrounds.
\begin{figure}[htbp]
	\centerline{\includegraphics[width=6.0in]{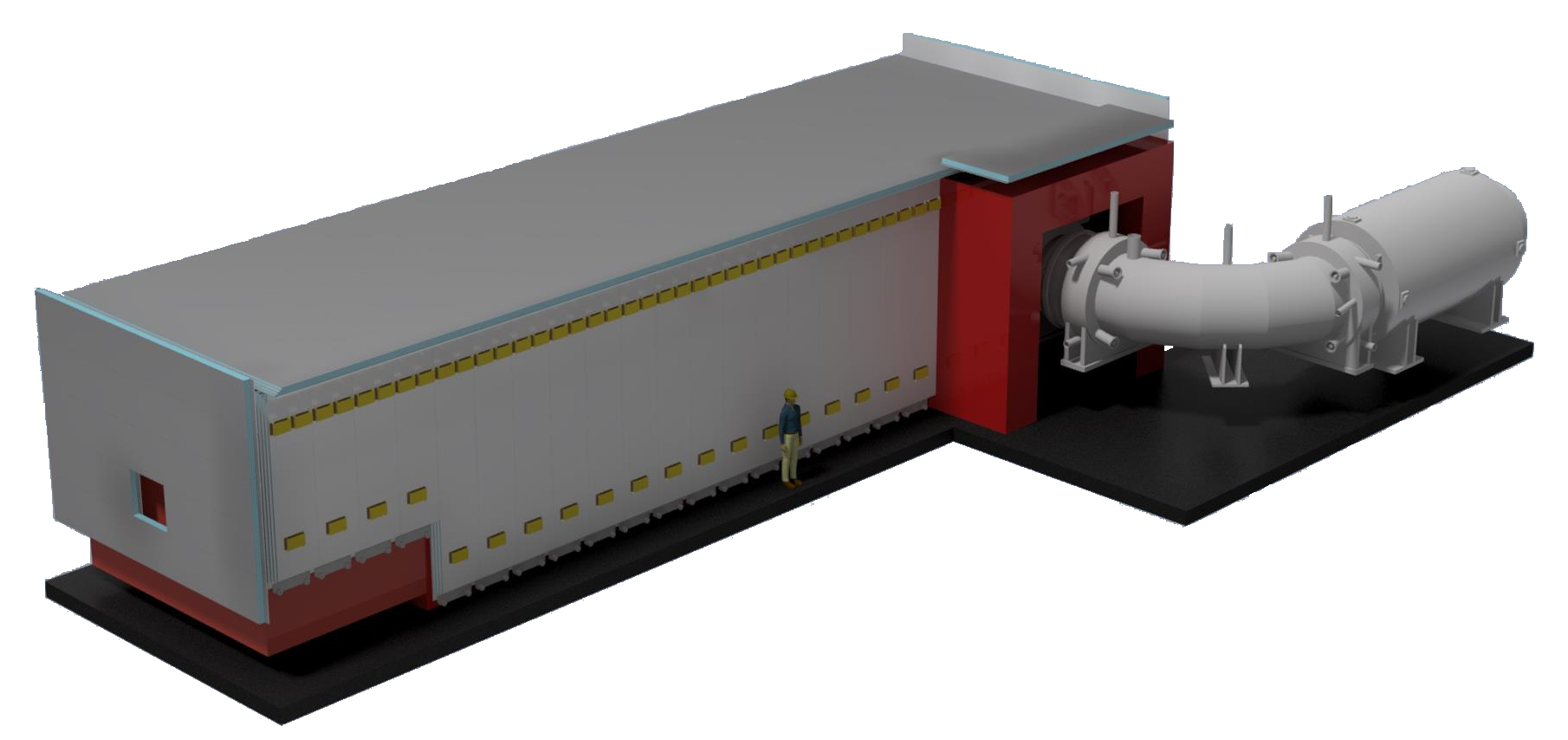}}
  \caption{
   In the image, concrete blocks (red) surround the Mu2e beamline which you see entering on the right. The cosmic ray veto detector is mounted on the top and four sides of the blocks (grey) and surrounds the Mu2e detector.}
  \label{fig:crv}
\end{figure}

Cosmic rays can produce conversion-like electrons
from muon decay-in-flight, neutron interactions, and muons misidentified as 
conversion electrons.  Such events are expected to appear at a rate of roughly 
one per day and represent the largest background to the experiment.  To identify 
such events, a cosmic ray veto (CRV) detector has 
been fabricated that will surround much of the Mu2e detector region (Fig.\,\ref{fig:crv}).  
The CRV employs silicon photomultipliers (SiPMs) whose performance is the subject
of this paper.  The system is designed to veto cosmic ray muons with a $99.99 \%$ efficiency, reducing the cosmic ray background expectation to significantly less than one event over the full experimental run. 

A calibration procedure for SiPMs is described here, including corrections for the temperature dependence of their light yield.  The SiPMs used in the CRV will not be kept at a constant temperature due to the complexity and cost of implementing such a system. Rather, the plan is to record the SiPM temperature, as well as the temperature of the device that provides the bias and amplifies the signal, and correct for the light yield dependence on these parameters. This manuscript describes how the SiPM temperature dependence was measured in a controlled environmental chamber.  How these measurements are used to calibrate the SiPM light yield response are described in detail below.  This note concludes by validating the calibration procedure on data from the temperature-controlled environmental chamber and from CRV modules stored at Fermilab.  The calibration steps and temperature measurements are described in detail to facilitate use by future SiPM-based detector systems that require equalized detector response as a function of temperature. 

\section{The Cosmic Ray Veto Detector Design}\label{sec:crv}
The CRV consists of four layers of scintillator bars with embedded wavelength-shifting fibers, read out by silicon photomultipliers. A conversion-like electron that comes in coincidence with a track stub found in the CRV is vetoed in the offline analysis.

The CRV scintillator bars are grouped in modules, each module consisting
of 64 bars, 16 per layer, with aluminum sheets of 3/8" (9.525\,mm) thickness 
placed between each scintillator layer. The entire CRV consists of 83 total modules.
Each scintillator bar is $51.34{\times}19.80$\,mm$^2$ in cross section, 
with lengths ranging from 0.985\,m to 6.900\,m.\footnote{The bars have a polystyrene 
base doped with 1\% PPO, and 0.03\% POPOP.}  The bars have a 
thin TiO$_2$ reflective coating and two channels into which the wavelength-shifting fibers are inserted~\footnote{Kuraray America, Inc.}. The fibers have diameters of 1.4\,mm and 1.8\,mm.

\begin{figure}[htbp]
\centering
\includegraphics[scale=0.4]{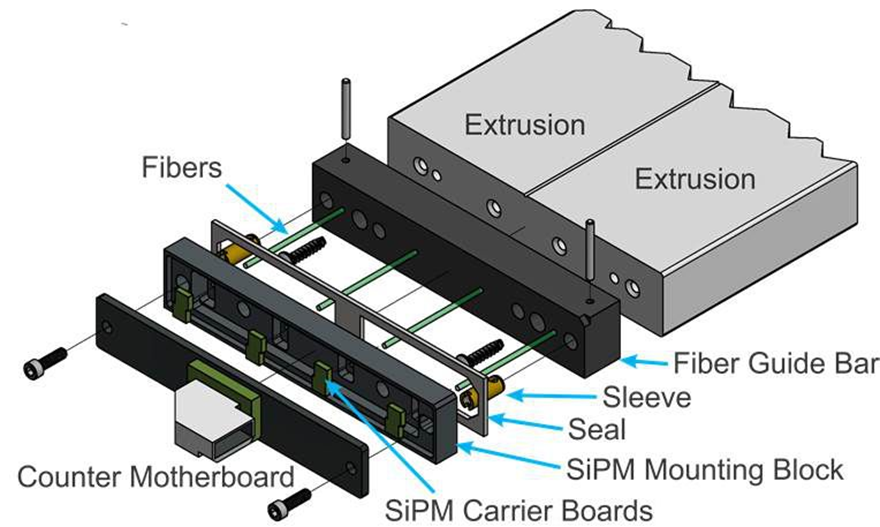}
\caption{Exploded view of the manifold at the end of a di-counter showing the fiber guide bar, SiPM mounting block, SiPM carrier boards, SiPMs, and counter motherboard. The flasher LEDs, thermometer, and pogo pins are not shown.\label{fig:dicounterreadout}}
\end{figure}

The fibers are read out by $2{\times}2$\,mm$^2$ silicon photomultipliers located at each end, except in special cases where only one side is utilized.~\footnote{The SiPMs are 
Hamamatsu S13360-2050VE devices with a 50\,$\mu$m pixel size, the large size being chosen as a compromise between dynamic range and quantum efficiency.}
For readout simplification, two bars are glued together side-by-side to form a di-counter (Fig.~\ref{fig:dicounterreadout}). An acetal fiber guide bar (FGB) is attached to each end of a di-counter, into which the fibers are glued with their ends polished. (The fibers are not glued into the scintillator channels.) Each SiPM is mounted on a printed circuit board by the manufacturer~\footnote{Hamamatusu Corp.} (called SiPM carrier boards, see Fig.~\ref{fig:dicounterreadout}). The SiPMs are housed in small rectangular wells in a black anodized aluminum fixture called the
SiPM mounting block, which is precisely registered to the FGB by a pair of
sleeves.  A small circuit board, the counter motherboard (CMB), has spring-loaded
``pogo" pins that gently press the SiPMs against the fibers and provide electrical 
contact to the SiPM.  It provides the SiPM bias voltage and sends the four SiPM output signals of each di-counter end
to a readout board via an HDMI cable.  The CMB also has two flasher LEDs, one for 
each bar, and a single thermometer centered on the board.\footnote{The thermometer is a UMW DS18B20U digital temperature sensor with a range from -55$^{\circ}$C to 125$^{\circ}$C and an accuracy of $\pm 0.4^{\circ}$C.}  This thermometer's temperature measurement is used as a proxy for the temperature of the four SiPMs serviced by the CMB.

The readout manifold, attached to each di-counter end, was designed to be as compact as possible to minimize
gaps in the coverage of the CRV that could allow cosmic-ray muons to leak through undetected.  Cooling the SiPMs to control their temperature,
and hence breakdown voltage, as well as to reduce their dark count rates, was
entertained, but rejected due to the strict gap requirements (needed to keep the
detection efficiency for cosmic-ray muons over 99.99\%) as well as cost considerations.
Rather, it was decided to carefully monitor the SiPM temperatures to allow for 
temperature corrections to be made to correctly determine the 
photoelectron (PE) yields.

All of the 83 modules that comprise the CRV have been fabricated and are
at Fermilab in the Wideband detector hall awaiting installation~\cite{Boi:2021yxw}.
The tests reported here come from several modules situated in a cosmic-ray
test stand, as well as from SiPMs installed inside and outside
of a temperature-controlled chamber also at Fermilab.

\section{Readout of the Cosmic Ray Veto}\label{sec:readout}

\begin{figure}[htbp]
\centering
\includegraphics[width=5in]{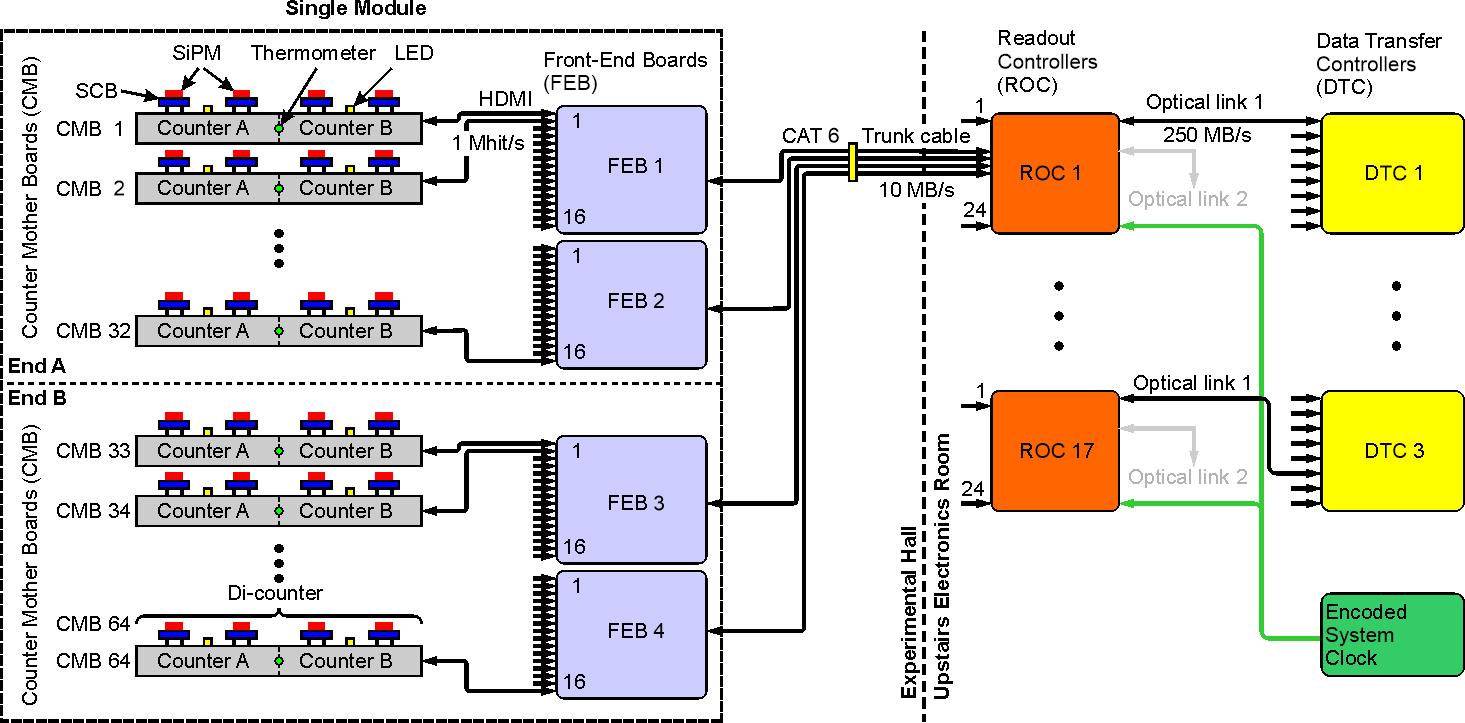}
\caption{Architecture of the CRV readout for a ``normal'' module read out on both ends. \label{fig:readout}}
\end{figure}

The architecture of the CRV readout is shown in Fig.\,\ref{fig:readout}.
Front-end boards (FEBs) provide the bias to the SiPMs and serve to take the signals 
from the SiPMs, amplify and shape them, digitize them in amplitude and time, 
zero-suppress them, and finally, store them for later readout.  Each FEB serves 64 SiPMs 
via 16 HDMI cables. Two FEBs are required to read out each end of a CRV module.

The FEBs receive power through CAT6 cables from a readout controller (ROC), 
which also furnishes the clock timing information.  In the normal readout of
the CRV, upon receiving a trigger, each ROC accumulates data from up to 24 FEBs 
and sends it, by means of optical links, to the Data Transfer Controllers (DTCs), which serve as the 
interface between the data acquisition hardware and the online computer farm. 

For the work described here, the DTCs were not used and the ROCs were only 
used to provide power, timing information, and triggers. The FEB data were read out 
via CAT6 cables, through a switch, to a DAQ PC. The PC was used to set the
FEB amplifier gain and SiPM bias voltages, as well as to retrieve the CMB 
temperatures and SiPM waveforms.

Each FEB has eight Analog Front End (AFE) chips\footnote{AFE5807, Texas Instruments, Inc., https://www.ti.com/product/AFE5807} that serve to amplify, shape,
and digitize (12-bit ADC) the signals from eight SiPMs at 80~MSPS 
(Mega samples per second or every 12.5\,ns). The digitized signals from the eight AFEs are sent to four FPGAs (each FPGA handling 16 channels). The FPGAs perform a set of functions --- such as zero-suppression --- that were not used in this work. The data from the
FPGAs are sent to a buffer memory where they await a trigger to be read out.

Each FEB has a temperature sensor that allows the temperature dependence
of the bias it provides to the SiPMs to be determined. The biases are
set by an overall bus bias common to all 64 SiPM channels served by one FEB, along with
trim settings that allow each SiPM's bias to be set to a precision of 2\,mV.

The data acquisition process at the Wideband hall consisted of two phases: a two-minute live spill period during which triggers are accepted by the FEB, followed by a one-minute off-spill period dedicated to transmitting the recorded data to the PC for storage. 
Before the onset of each spill, slow control information was retrieved from the FEBs and written to the raw data file. This information included the settings of the AFEs, bias settings for each SiPM, CMB and FEB temperature, and the AFE gain settings.

\begin{figure}
    \centering
    \includegraphics[width=0.85\linewidth]{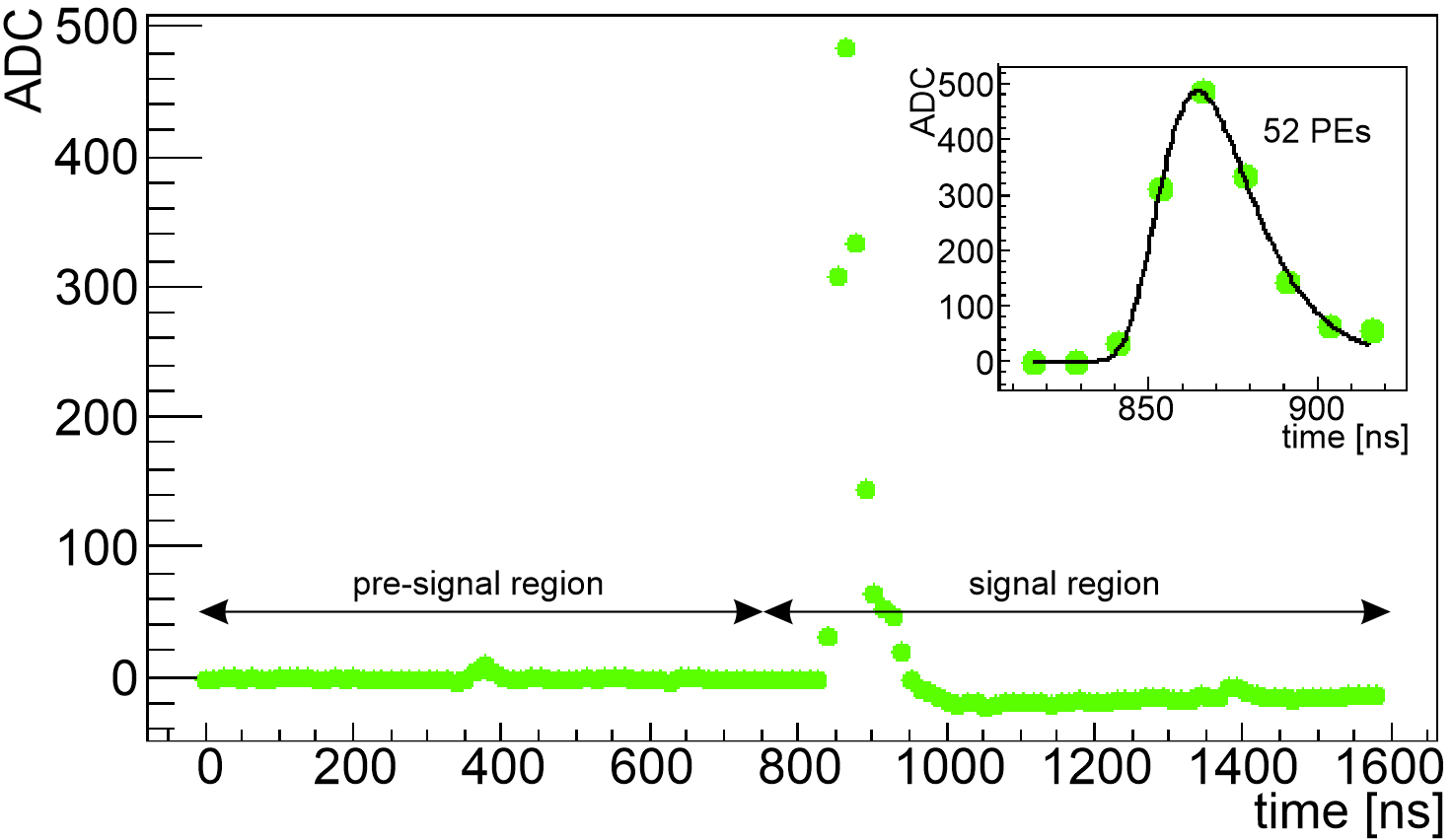}
\caption{A typical waveform and noise pulse appearing in a 127 ADC sample time slice, each sample separated by 12.5\,ns. The pre-signal region shows a dark-count pulse of 1\,PE at 400\,ns. The signal region shows a signal pulse with 52\,PEs. All pulses get fitted with a Gumbel function.}
\label{fig:waveform}
\end{figure}

During the live spill, the digitized SiPM-generated data were continuously written to the 
buffer memory on the FEB (at 80\,MSPS). Upon receiving a trigger, data before 
and after the trigger time were transmitted. Typically 750\,ns of data were read out
before the trigger (pre-signal region --- see Fig.\,\ref{fig:waveform}), allowing the pedestal to be determined, and capturing an occasional dark count from the SiPM noise, the latter being used in the calibration process.
A total of 127 ADC samples were recorded for each of the 64 FEB channels. 
This sequence of data initiated by a single trigger is called an event. 
The event rate was approximately 3\,Hz for the cosmic-ray trigger described below.
For the LED flasher trigger, also described below, the event rate was 1 kHz. 

\section{Calibration}\label{sec:calibration}

\subsection{Calibration Overview at a Stable Temperature}
\label{sec:calibration_no_T}
In the pre-signal region, the data in each channel undergoes a scan for dark pulses. An example dark pulse can be seen in Fig.\,\ref{fig:waveform}. The pulses undergo pedestal subtraction and are fitted using a Gumbel function given by the equation:

\begin{equation}\label{eq:gumbel}
f(t) = A \ e^{\frac{t-\mu}{\beta}-e^{\frac{t-\mu}{\beta}}},
\end{equation}
where $\mu$ is the time of the pulse peak, $\frac{A}{e}$ is the pulse height ($e$ is Euler's number), 
and $A \cdot \beta$ is the pulse area. Note that the Gumbel function \cite{gumbel} was selected empirically based on the quality of pulse fitting.  

\begin{figure}
    \centering
    \includegraphics[width=0.6\linewidth]{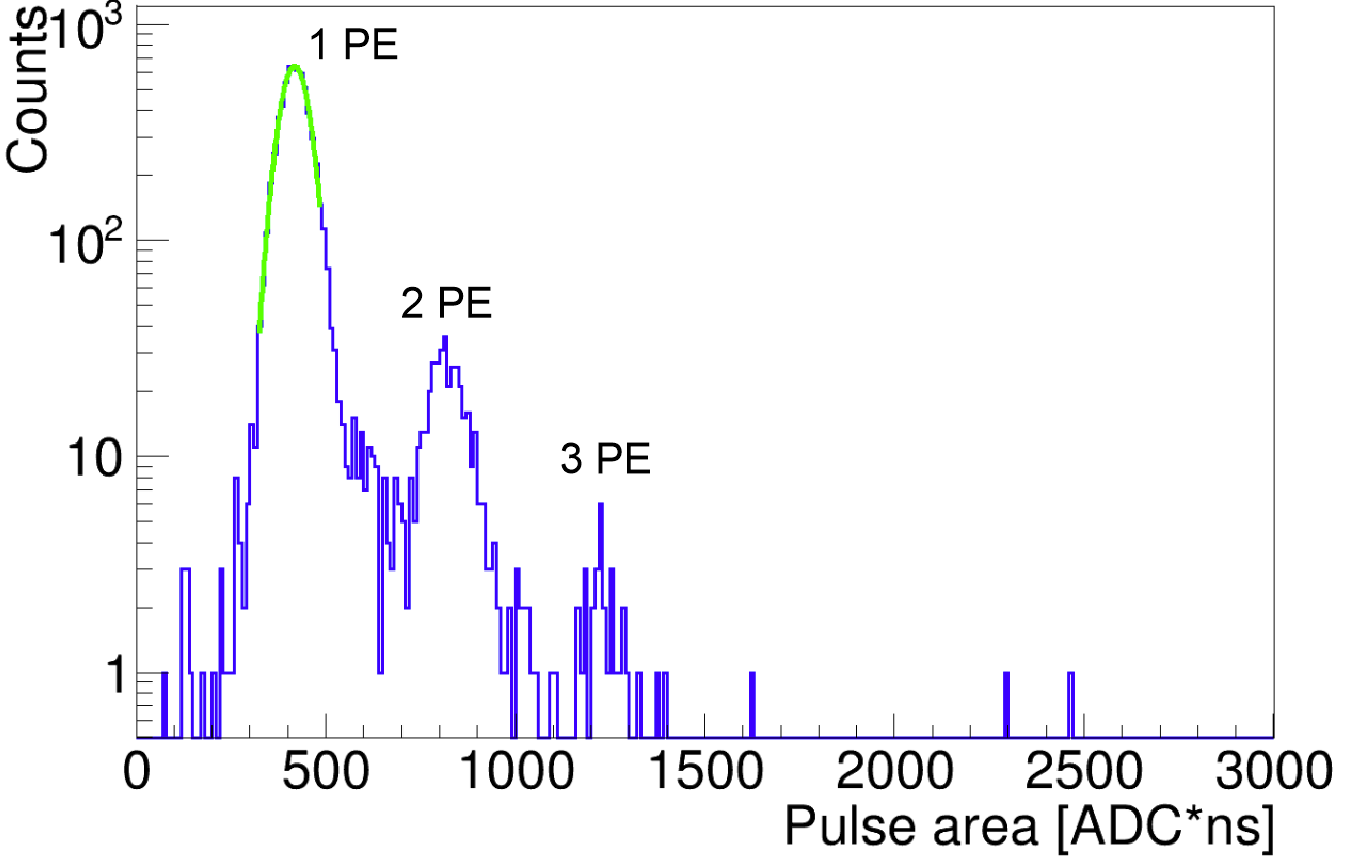}
\caption{The pedestal-subtracted dark count distribution of a single SiPM displays distinct peaks that correspond 
to one, two, and three photoelectrons. The location of the first peak (the single photoelectron peak, SPE) is 
utilized to extract the SPE calibration constant.}
\label{fig:calibration}
\end{figure}

\begin{figure}
  \begin{subfigure}{0.47\textwidth}
    \includegraphics[width=\linewidth]{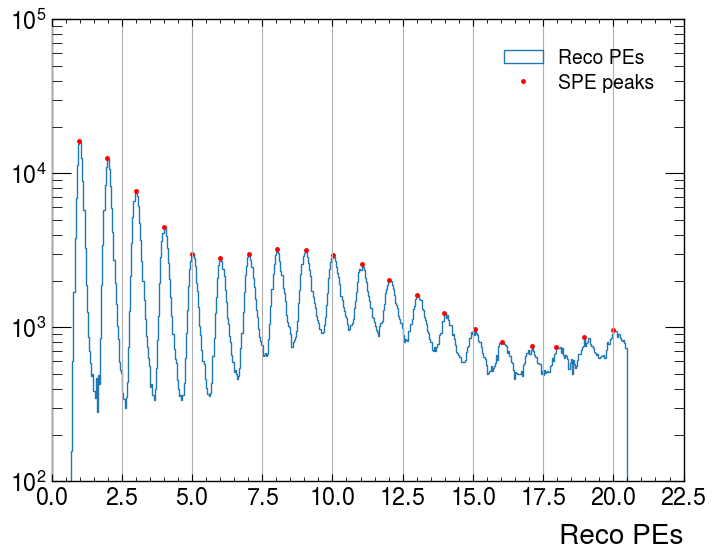}
  \end{subfigure}
  \begin{subfigure}{0.5\textwidth}
    \includegraphics[width=\linewidth]{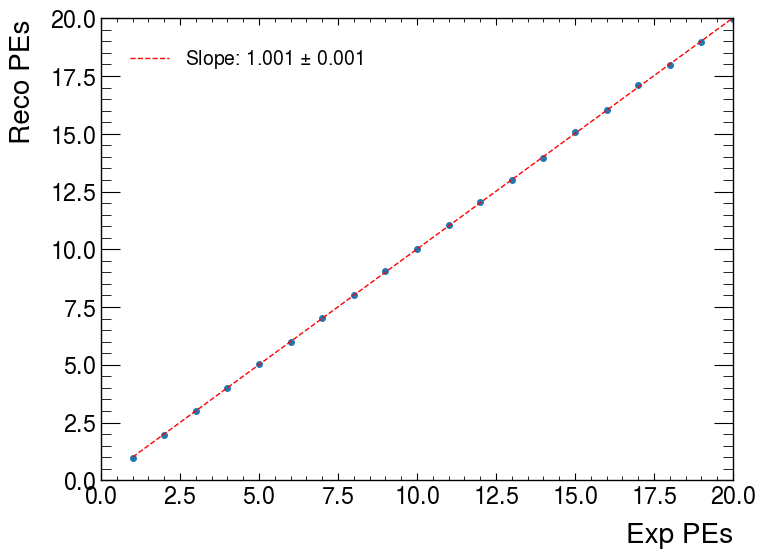}
  \end{subfigure}
\caption{Left figure: the distribution of the integrated photoelectron peaks (from a single SiPM channel) collected
when the channel is illuminated with the LED flasher.  Right figure: the integrated photoelectron peaks versus peak number showing a linear trend with the slope of $1.001{\pm}0.001$.  Note that these
peaks have not been corrected, and were taken at the same temperature.}
\label{fig:led_runs}
\end{figure}

The fitting process determines the dark count pulse areas, the distribution of which can be seen in 
Fig.\,\ref{fig:calibration}, showing distinct peaks corresponding to the pulse areas of 1 PE, 
2 PEs, and so forth.\footnote{Note that strictly speaking SiPMs produce a signal that is 
proportional to the number of hit pixels, not PEs.  At low light levels the correspondence
between the two is quite good.  In this paper, we will use PEs rather than hit pixels.} 
The single photoelectron peak (SPE) in the distribution of Fig.\,\ref{fig:calibration} is used 
to produce the SPE calibration constant that translates the SiPM waveform area into the reconstructed 
number of photoelectrons.\footnote{Due to a low cross-talk rate and hence limited statistics 
for the photoelectron peaks greater than 1 PE, only the SPE peak is utilized to establish the SPE
calibration constant.} The stability of the SPE calibration constant across a broader range of PEs is verified through 
LED flasher runs, where the LED flashers can be tuned in such a way that they produce only low PE pulses. For these kinds of signals, individual PE peaks can be easily separated as shown in  Fig.\,\ref{fig:led_runs}, where Reco PEs stands for peaks in
the SiPM output as fitted by the Gumbel function.\footnote{When determining pedestals we have subtracted 0.13 ADC counts from their mean values in order to get a more perfect linear fit between the reconstructed PEs and the expected PEs.}

The calibration procedure described above works well for reasonably stable temperature environments and has been used in test beam experiments to study the CRV performance~\cite{Mu2e:2017lae}.  However, when efforts to precisely study the CRV aging properties over several years were attempted, it became clear that the temperature dependence of the CRV performance could not be ignored.\footnote{With the calibration scheme described here, precise aging measurements of the CRV counters are possible in the non-temperature controlled environment in which the counters are stored; these will be reported in a forthcoming publication.}  Below we describe the updated calibration procedure for equalizing the CRV response to a temperature that was primarily derived to make aging measurements of the CRV counters, but will also be valuable in the operation of the Mu2e experiment.    

\subsection{Calibrating for the  Temperature Dependence}
\begin{figure}[ht]
    \centering
    \includegraphics[width=1\linewidth]{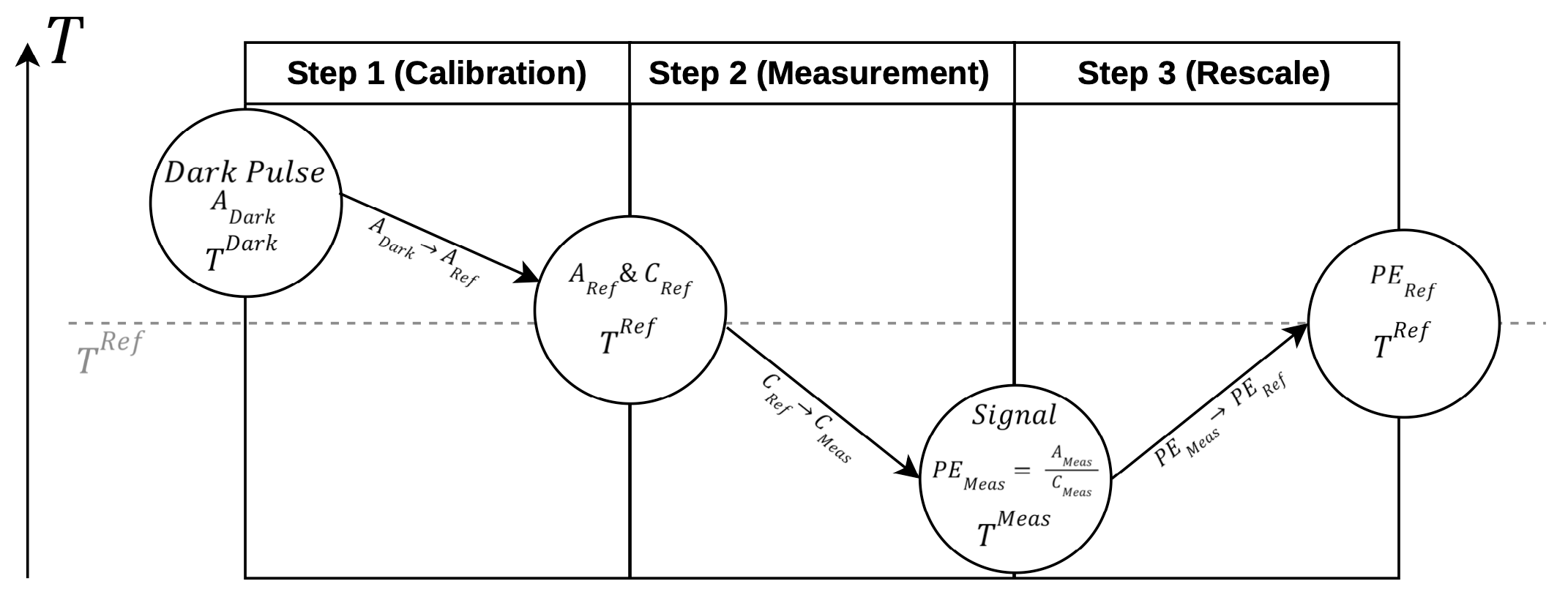}
\caption{A flow diagram of the calibration method to equalize the CRV response as a function of temperature.  The calibration, measurement, and rescale steps are described in detail in the text.  Note that it takes many dark pulses to derive the SPE calibration constant.  Each dark pulse must have its response, taken at temperature, $T_{\rm Dark}$,  corrected to the reference temperature, $T_{\rm Ref}$. Similarly, $A_{\rm Meas}$ is a signal pulse that can be taken at any temperature, $T_{\rm Meas}$.  Before the SPE calibration constant can be applied it must be corrected from the reference value, $C_{\rm Ref}$, to the relevant value at the measured temperature, $C_{\rm Meas}$. The parameter definitions required for the temperature corrections are given in Table\,\ref{tab:corrections}, and their measurement is described in Sec.\,\ref{sec:parameters}. }
\label{fig:flow}
\end{figure}

The SPE calibration constants and photo-detection efficiencies are impacted by the temperatures of both 
the SiPMs (as measured by the thermometer on the CMBs) and FEB ($T_{\rm SiPM}$ and $T_{\rm FEB}$), the former because the SiPMs have temperature
dependent breakdown voltage, and the latter because of the bias provided by the FEBs to the SiPMs
is slightly temperature dependent and therefore not exactly at its set value. In addition, the gain of the AFE 
amplifier/digitizers have a temperature dependence.
Consequently, a temperature correction is essential to 
adjust the SPE constants and photo-detection efficiencies to reference temperatures, chosen
to be: $T_{\rm SiPM}^{\rm Ref}=$ 20$^{\circ}$C for the SiPM and $T_{\rm FEB}^{\rm Ref}=$ 40$^{\circ}$C for the FEB
(roughly, their usual operating temperatures). 
This adjustment follows a specific procedure, as outlined in Fig.\,\ref{fig:flow} and described in detail below.

In Step 1, described in Sec.\,\ref{sec:ref_cal}, dark pulses are collected, fitted, and the pulse areas are extracted, $A_{\rm Dark}$.  
The SiPMs are not noisy, and hence the collection of sufficient numbers of dark pulses for a precise measurement of the SPE calibration constant might occur on a time scale in which the temperature is not constant.  Hence, the dark pulses must first be corrected to the reference temperature before the SPE calibration constant, $C_{\rm Ref}$, is extracted from the single photoelectron peak.  
In Step 2, described in Sec.\,\ref{sec:reconstruction}, the calibrations are applied to a measured signal pulse area, $A_{\rm Meas}$, at an arbitrary temperature, $T^{\rm Meas}$.  $C_{\rm Ref}$ must first be corrected to the proper value at the measurement temperature, $C_{\rm Meas}$.  It can then be applied to the measured pulse area, $A_{\rm Meas}$, to estimate the number of photoelectrons observed.  To compare the performance of the CRV over time, Step 3 of Fig.\,\ref{fig:flow}, also described in Sec.\,\ref{sec:reconstruction}, can be applied to rescale the measured PE value to the reference temperature.  In this way, CRV performance over time can be compared even if the temperature is not stable.  

\begin{table}[ht]
  \centering
  \begin{tabular}{cp{3in}c} 
    \hline\hline
\multicolumn{2}{c}{Parameter} & Value \\ [0.5ex] 
    \hline
 $p_1$ & Temperature coefficient of the SiPM breakdown voltage                & 55.4${\pm}$0.8 mV/K \\ 
 $p_2$ & Temperature coefficient of the SiPM bias provided by the FEB         & 4.1$\pm$0.7 mV/K \\
 $p_3$ & SPE calibration constant (or pulse area) increase per SiPM bias increase   & 125$\pm$3 ADC${\cdot}$ns/V \\
 $p_4$ & SPE calibration constant (or pulse area) decrease due to AFE temperature dependence  & -1.46$\pm$0.16 ADC$\cdot$ns/K \\
 $p_5$ & PE increase due to overvoltage bias increase                         & 21.8$\pm$2.1\%/V \\ [1ex] 
 \hline\hline
  \end{tabular}
  \caption{Temperature correction parameters extracted from the temperature-dependence studies described in Sec.~\ref{sec:parameters}. The correction parameters $p_3$, $p_4$, and $p_5$ are valid for a particular gain setting and a nominal over-voltage of 2.5\,V. }
  \label{tab:corrections}
\end{table}

The CRV SiPMs are operated at constant bias voltages, corresponding to about 2.5~V over breakdown calculated at the reference temperature. To achieve this, we have developed a bias scan procedure to extract the breakdown voltages for each SiPM. These breakdown voltages depend on the SiPM temperatures ($p_1$ in Table\,\ref{tab:corrections}). Furthermore, temperature variations in FEBs result in relatively small temperature-dependent fluctuations in the bias voltage supplied to SiPMs ($p_2$ in Table\,\ref{tab:corrections}). 

Changes in over-voltage due to temperature fluctuations, in turn, lead to corrections of the SPE constants ($p_3$ in Table\,\ref{tab:corrections}). The SPE constants are also influenced by a gain change in the AFE ($p_4$ in Table\,\ref{tab:corrections}).\footnote{Parameters $p_3$ and $p_4$ depend on the AFE gain settings, and the values in the table are only valid for the specific experimental setup.}  Finally, changes to the over-voltage also require corrections to the light yield ($p_5$ in Table\,\ref{tab:corrections}).

\subsection{Obtaining the Reference Calibration}
\label{sec:ref_cal}

Each pulse collected during the dark count scan undergoes an adjustment of its measured pulse area $A_{\rm Dark}$ to its value at the reference temperatures $T_{\rm SIPM}^{\rm Ref}$ and $T_{\rm FEB}^{\rm Ref}$. For each pulse, the deviation of the over-voltage $\Delta{V_{\rm ov}}$ due to the temperature difference is calculated:
\begin{equation}
\Delta{ V_{\rm ov}} = - p_1 \cdot ({\rm T}_{\rm SiPM}^{\rm Dark} - {\rm T}_{\rm SiPM}^{\rm Ref}) + p_2 \cdot ({\rm T}_{\rm FEB}^{\rm Dark} - {\rm T}_{\rm FEB}^{\rm Ref}).
\label{eq:calibOvervoltageChange}
\end{equation}
This deviation of the over-voltage is used to calculate what the pulse area at the reference temperatures of the CMB and FEB would have been. The gain change due to the FEB temperature is an additional adjustment:
\begin{equation}
A_{\rm Ref} = A_{\rm Dark} - p_3 \cdot \Delta{V_{\rm ov}} -  p_4 \cdot (T_{\rm FEB}^{\rm Dark} - T_{\rm FEB}^{\rm Ref}).
\label{eq:calibPulseareaChange}
\end{equation}
These temperature-corrected pulse areas $A_{\rm Ref}$ are then histogrammed in a calibration plot, and the temperature-corrected SPE constants, $C_{\rm Ref}$, are determined using the same approach as described in Sec.~\ref{sec:calibration_no_T} and shown in Fig.~\ref{fig:calibration}. The parameters in Eq.~\ref{eq:calibPulseareaChange} were obtained for temperature corrections of the SPE calibration constant, which is the pulse area of a single photoelectron. Therefore, this equation is only valid for pulse areas of single PEs, and only the first PE peak of the temperature-corrected calibration plot is used to determine the SPE calibration constant.

\section{Reconstructing the Light Yield}
\label{sec:reconstruction}

SiPM signal pulses are pedestal-subtracted and fitted with a Gumbel function, shown in Eq.~\ref{eq:gumbel}, in the same way as the dark counts.\footnote{Note that the temperature dependence of the pedestals were found to be negligible.} Their pulse areas, peak times, leading-edge times, and width parameters are recorded. 

\begin{figure}
  \centering
  \begin{subfigure}[b]{0.49\textwidth}
    \includegraphics[width=\linewidth]{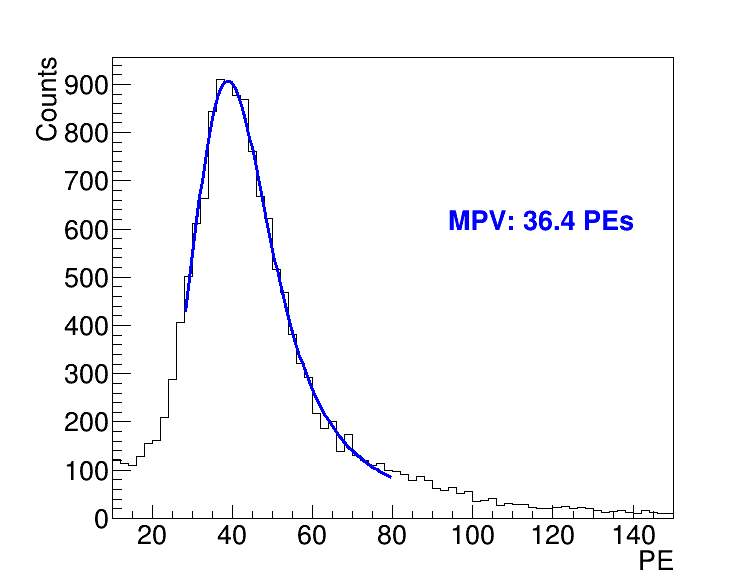}
    \label{subfig:mip}
  \end{subfigure}
  \begin{subfigure}[b]{0.49\textwidth}
    \includegraphics[width=\linewidth]{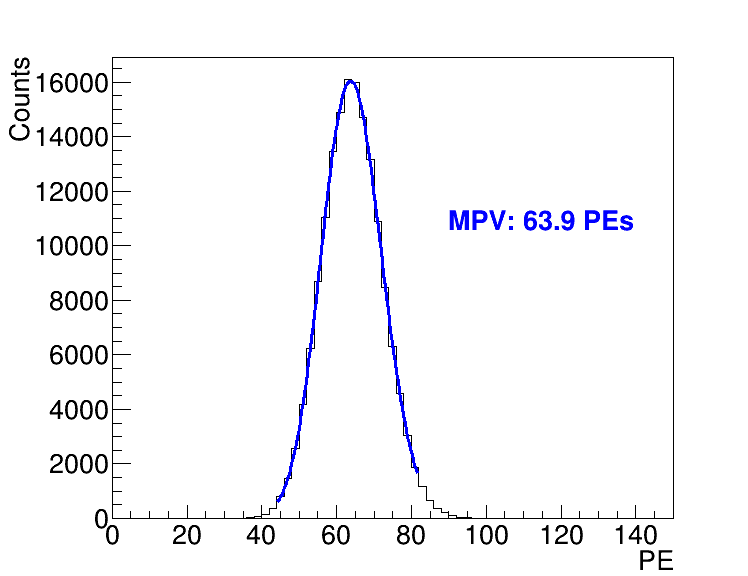}
    \label{subfig:led}
  \end{subfigure}
  \caption{Left figure: typical SiPM photoelectron distribution example from cosmic-ray muons, fitted with a convoluted Gauss-Landau function. The plot is zero-suppressed to prevent it from being dominated by low-PE dark counts. Right figure: SiPM response to the LED flasher, fitted with a Poisson function. The PE values used in both histograms were adjusted to the reference temperatures using the temperature-correction procedure outlined in the text.}
  \label{fig:pe}
\end{figure}

Similar to the SPE calibration constants $C_{\rm Ref}$, the measured pulse requires temperature corrections. The process begins by obtaining the SPE calibration constant for the temperatures at the time the pulse was recorded, $C_{\rm Meas}$. Since the SPE calibration constant $C_{\rm Ref}$ was initially stored for the reference temperatures $T_{\rm SiPM}^{\rm Ref}$ and $T_{\rm FEB}^{\rm Ref}$, the temperature-correction procedure outlined in the previous section must be reversed to derive the SPE calibration constant specific to the SiPM and FEB temperatures of the given pulse $T_{\rm SiPM}^{\rm Meas}$ and $T_{\rm FEB}^{\rm Meas}$. As before, the first step is calculating the deviation of the over-voltage $\Delta{V_{ov}}$ due to the temperatures at the SiPM and FEB (see Eq.~\ref{eq:calibOvervoltageChange}):
\begin{equation}
\Delta{V_{\rm ov}} = - p_1 \cdot (T_{\rm SiPM}^{\rm Meas} - T_{\rm SiPM}^{\rm Ref}) + p_2 \cdot (T_{\rm FEB}^{\rm Meas} - T_{\rm FEB}^{\rm Ref}).
\label{eq:recoOvervoltageChange}
\end{equation}
The second step is adjusting the calibration SPE constant from its value at the reference temperatures ($C_{\rm Ref}$) to a calibration constant at the actual temperatures when the pulse was recorded ($C_{\rm Meas}$). This is done using the over-voltage change from Eq.~\ref{eq:recoOvervoltageChange} and the gain change due to the FEB temperature as an additional adjustment (analogous to Eq.~\ref{eq:calibPulseareaChange} above but adjusting the calibration constant instead of the pulse area):
\begin{equation}
C_{\rm Meas} = C_{\rm Ref} + p_3 \cdot \Delta{V_{\rm ov}} +  p_4 \cdot (T_{\rm FEB}^{\rm Meas} - T_{\rm FEB}^{\rm Ref}).
\label{eq:calibChange}
\end{equation}
Dividing the pulse area $A$ by this calibration constant $C_{\rm Meas}$ results in the number of photoelectrons of this pulse:
\begin{equation}
{PE_{\rm Meas}} = A_{\rm Meas} / C_{\rm Meas}.
\end{equation}\label{eq:recoCalibration}
The change in over-voltage affects the photon detection efficiency. To obtain the PE yield at the reference temperature, $PE_{\rm Ref}$, the effect of the over-voltage change has to be removed in a second step involving $p_5$ from Table~\ref{tab:corrections} and $\Delta{V_{\rm ov}}$ from Eq.~\ref{eq:recoOvervoltageChange}:
\begin{equation}
{PE_{\rm Ref}} = \frac{PE_{\rm Meas}}{1+p_5 \cdot \Delta{V_{\rm ov}}}.
\end{equation}\label{eq:recoCorrected}
The temperature-corrected number of photoelectrons, ${PE_{\rm Ref}}$, are accumulated in a histogram (an example is shown in Fig.~\ref{fig:pe}). The resultant PE distribution is fitted with a convoluted Gauss-Landau function for cosmic-ray data and a Poisson function for LED data. This fit determines the most probable value of the PE distribution and is a useful figure of merit to characterize the CRV performance.

\section{Determination of Temperature Correction Parameters} \label{sec:parameters}

\begin{figure}
\centering
\includegraphics[width=0.95\textwidth]{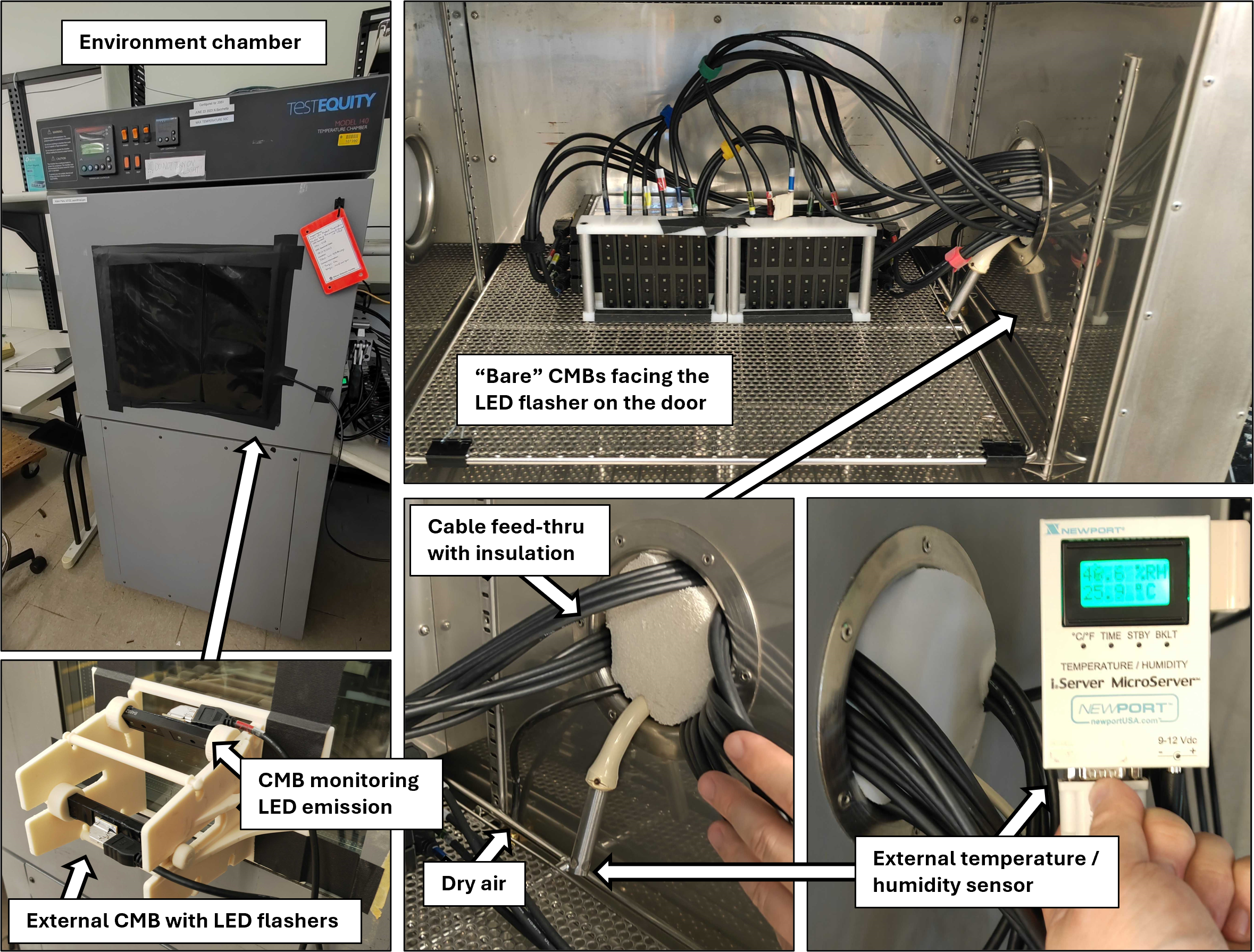}
\caption{Photographs of the SiDet setup used to study temperature dependencies. \label{fig:sidetpictures}}
\end{figure}

To determine the temperature corrections, LED data were collected and studied at Fermilab's SiDet facility using the environmental chamber illustrated in Fig.~\ref{fig:sidetpictures}. We measured the temperature dependence of the SiPMs, with the bias provided by a FEB situated outside of the chamber. During the studies described in this section, the AFE chip gains were uniformly set to their default operating value.\footnote{The AFE gain was set to 33\,dB, for a voltage gain of 45.} The SPE-related parameters would change should the AFE chip gain vary.

\subsection{SiPM Temperature Corrections}\label{sec:sipmtemp}

\begin{figure}
\centering
\includegraphics[width=0.8\textwidth]{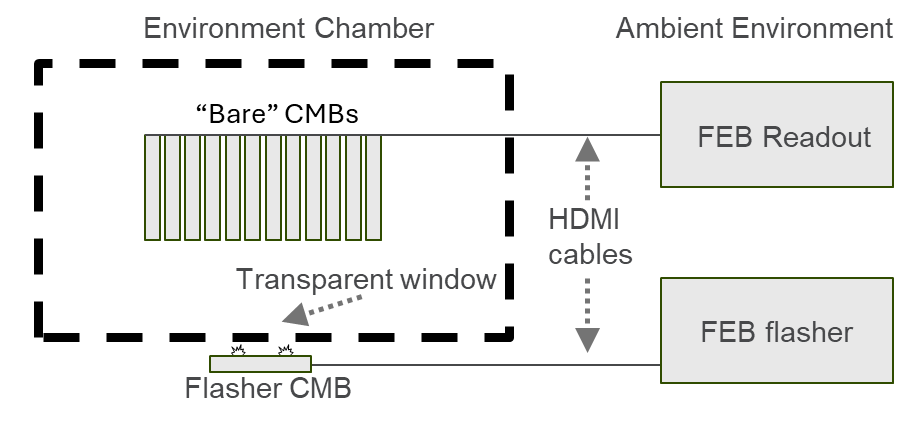}
\caption{Diagram of the setup for the SiPM temperature correction studies at SiDet. An array of 64 SiPMs, serviced by 16 CMBs not attached to any counters, and hence ``bare'', are situated inside the environmental chamber. A CMB is used to flash light through a window to the SiPMs inside the chamber. Not shown is an additional CMB outside the chamber with SiPMs biased to a constant voltage that is used to monitor the LED light emission.}
\label{fig:sidetdiagram}
\end{figure}

The configuration illustrated in Fig.~\ref{fig:sidetdiagram} was used to measure the temperature dependence of the SiPMs. Bare SiPMs (uncoupled to fibers) were placed inside the environmental chamber, each group of four serviced by one CMB.  The average of the CMB sensor readings was used as a proxy for the SiPM temperatures, and was found to agree well with the set temperature of the chamber.  An outside CMB flashed LED light through a transparent window. Two FEBs were employed in making the measurements. One was used to provide the SiPM biases and read out the SiPM signals; the other was used to flash the LED mounted on the outside CMB.  Both were exposed to a relatively constant room temperature. The SiPMs' dark pulses and their responses to the LED flashes were recorded for the chamber temperature range from -10$^{\circ}$C to 40$^{\circ}$C. For each temperature setting, the SiPM response was taken at a biases between 2\,V and 6\,V over-voltage (as calculated from the datasheet values provided by Hamamatsu). The applied bias voltages were recorded as $V_{\mathrm{bias}}$.

\subsubsection{SiPM Breakdown Voltage and Temperature Dependence} \label{sec:sipmvbdtemp}

While the SiPM breakdown voltage ($V_{\rm bd}$) is often determined by the current-voltage curve (``IV curve") of a device \cite{Nagy:2016mfv}, we opted to determine it using the gain vs bias technique using the single photoelectron peaks.

\begin{figure}
  \centering
  \begin{subfigure}[b]{0.49\textwidth}
    \includegraphics[width=0.9\linewidth]{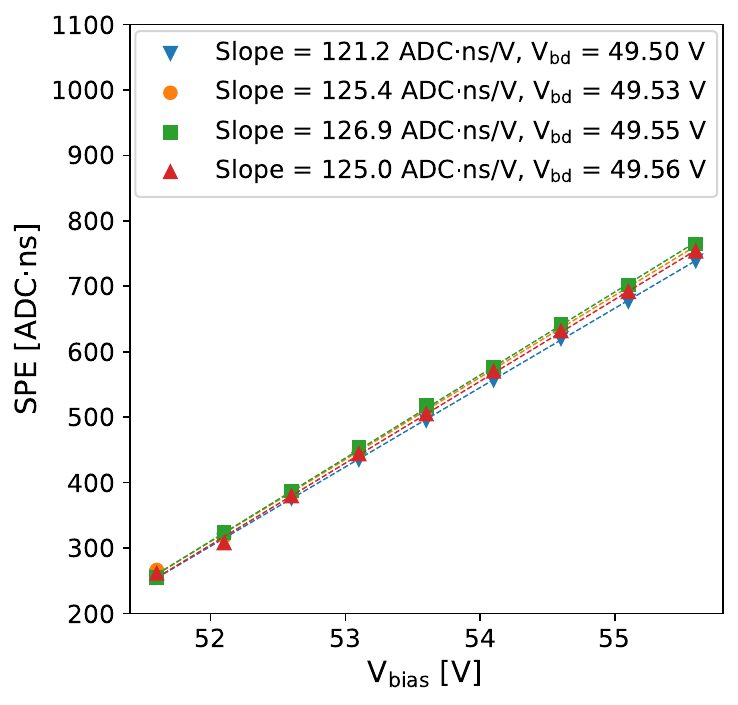}
    \label{subfig:mip2}
  \end{subfigure}
  \begin{subfigure}[b]{0.49\textwidth}
    \includegraphics[width=0.9\linewidth]{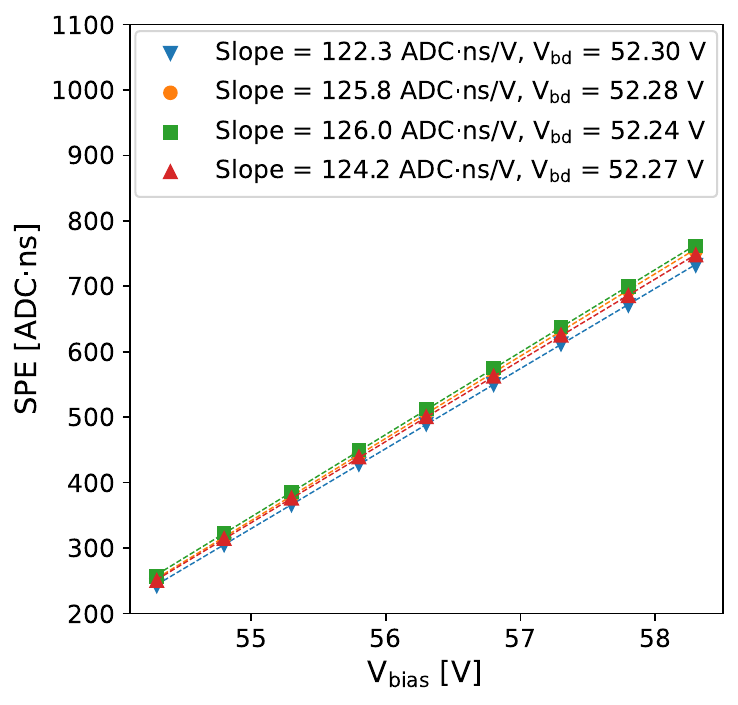}
    \label{subfig:led2}
  \end{subfigure}
  \caption{The SiPM SPE response versus bias voltage for four example SiPMs in the environmental chamber taken at -10°C (left) and 40°C (right). Extrapolation to SPE = 0 yields the SiPM breakdown voltages $V_{\rm bd}$ shown in the legend and plotted for all SiPMs and temperatures in Fig.~\ref{fig:Vbd_bychannel}.}
  \label{fig:SPE_bias_scan}
\end{figure}

\begin{figure}
  \centering
  \includegraphics[width=0.75\textwidth]{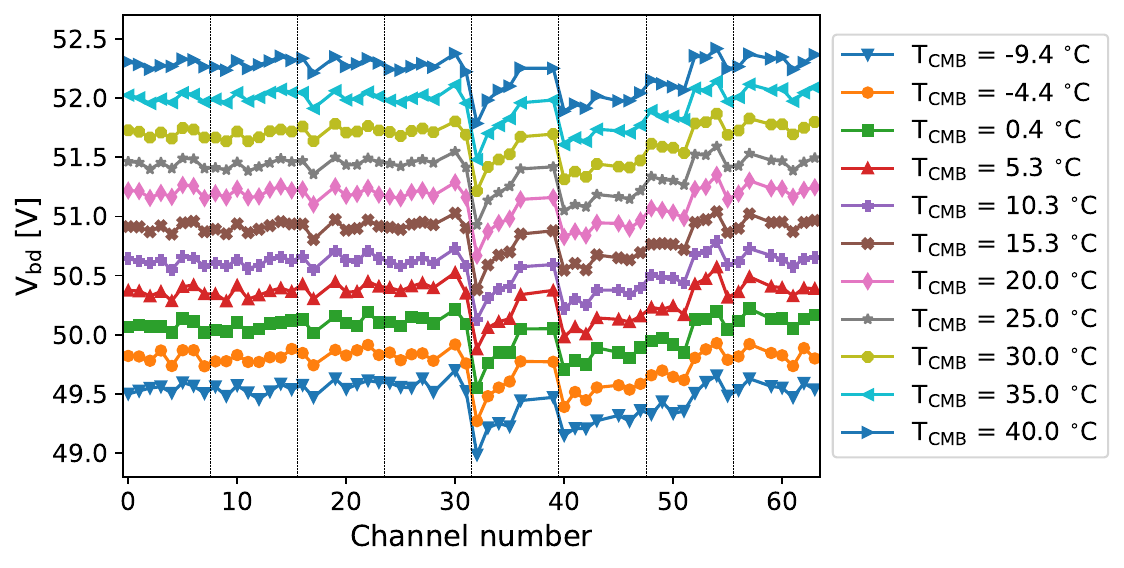}
  \caption{Breakdown voltages at each temperature for all 58 SiPMs.}
  \label{fig:Vbd_bychannel}
\end{figure}

Bias scans were performed at each temperature setting of the environmental chamber, and the SPE response was recorded.  The results, given in Fig.~\ref{fig:SPE_bias_scan}, show a linear dependence on the bias. The breakdown voltage $V_{\rm bd}$ is determined by extrapolating the linear fits to the point where SPE~=~0. The extracted breakdown voltages for all 58 SiPMs and each temperature are shown in the top plot of Fig.~\ref{fig:Vbd_bychannel}.\footnote{Fluctuations in the breakdown voltages of SiPMs 30--40 are attributed to production effects.  Note that these same variations were observed in independent quality control measurements of the SiPMs.  The SiPMs used in these studies were selected randomly from several batches received from the manufacturer and some variation is not surprising.}
Note that although 64 SiPMs were inside the chamber, 6 of them has issues that prevented their proper readout and were not used in the analysis. The four SiPMs shown as examples in Fig.~\ref{fig:SPE_bias_scan} were selected arbitrarily; the same devices are used in subsequent figures to illustrate example results.

\begin{figure}
  \centering
  \begin{subfigure}[b]{0.49\textwidth}
    \includegraphics[width=0.92\linewidth]{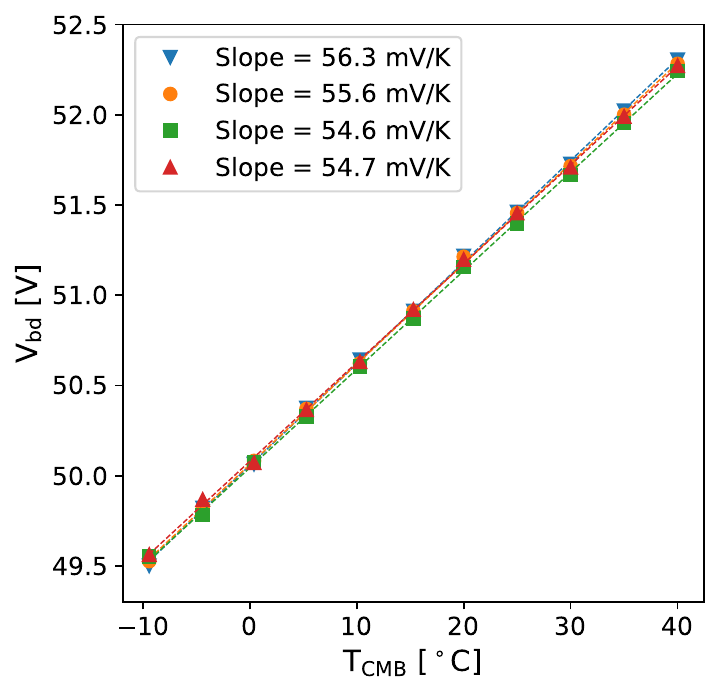}
    \label{subfig:mip3}
  \end{subfigure}
  \begin{subfigure}[b]{0.49\textwidth}
    \includegraphics[width=0.9\linewidth]{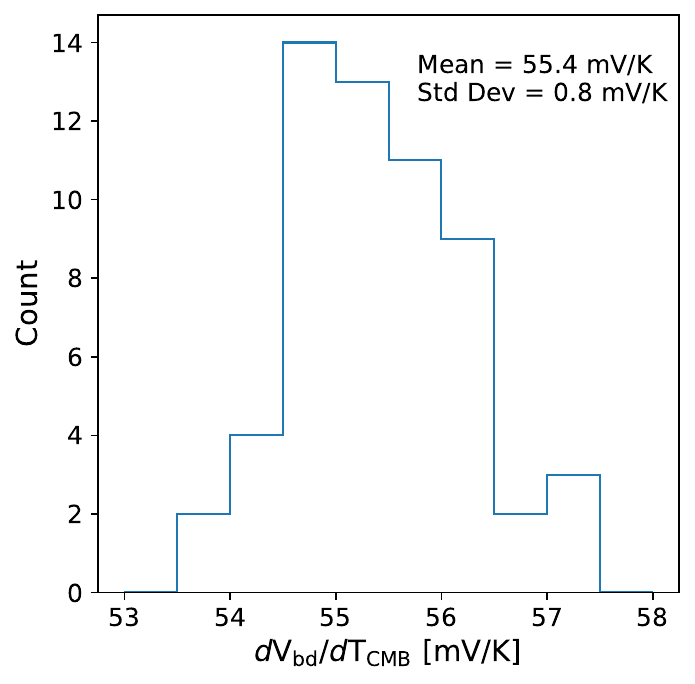}
    \label{subfig:led3}
  \end{subfigure}
    \caption{Left: Temperature dependence of the extracted breakdown voltage for the example SiPMs. The slope of the fitted line represents the breakdown voltage temperature coefficient. Right: Histogram of the temperature coefficients for the 58 SiPMs in our setup.}
  \label{fig:SPE_bias_scan_Vbd}
\end{figure}

The well-known increase in breakdown voltage with increasing temperature is apparent from Fig.~\ref{fig:Vbd_bychannel}.  Averaging over all SiPMs, the extracted breakdown voltage increases at the rate of $55.4{\pm}0.8$\,mV/K (Fig.~\ref{fig:SPE_bias_scan_Vbd} right). This is parameter $p_1$ in Table~\ref{tab:corrections}. It is worth noting that this value agrees with the specifications of 54\,mV/K provided by Hamamatsu. 


\subsubsection{SiPM Single Photoelectron Yield Dependence on Over-Voltage}\label{sec:vbdvover}
\begin{figure}
\centering
\includegraphics[width=0.75\textwidth]{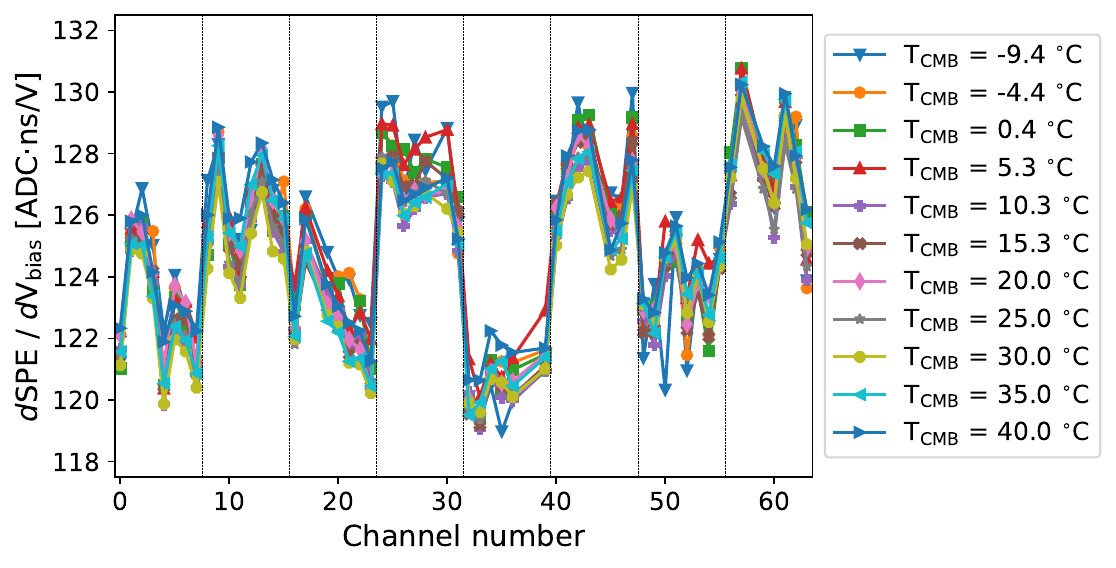}
\caption{The SPE increase per volt at each temperature for each SiPM, as extracted from the slopes of the linear fits, an example of which is  shown in Fig.~\ref{fig:SPE_bias_scan}.}
\label{fig:SPEperV}
\end{figure}

The temperature dependence of the breakdown voltage affects the photoelectron yield of the SiPMs, and hence the SPE calibration constants. That dependence was measured to be  $125{\pm}3\,\mathrm{ADC \cdot ns/V}$, from the fit parameters given in Fig.~\ref{fig:SPE_bias_scan}.  This is parameter $p_3$ in Table~\ref{tab:corrections}.  It is specific to the AFE chip gain setting we use for the CRV. The rate exhibits variations due to SiPM and AFE chip gain differences, but notably, the rate of SPE increase remains independent of the operating temperature over the temperature range tested (Fig.~\ref{fig:SPEperV}).  Since the SiPM-to-SiPM variation is small, using the average value of $125~\mathrm{ADC \cdot ns/V}$ yields satisfactory results, as will be shown in Sec.~\ref{sec:tempcorrectionsanitychecks} and Sec.~\ref{sec:validation}.

\subsubsection{SiPM Light Yield Dependence on Over-Voltage}

\begin{figure}
  \centering
  \begin{subfigure}[b]{0.49\textwidth}
    \includegraphics[width=0.9\linewidth]{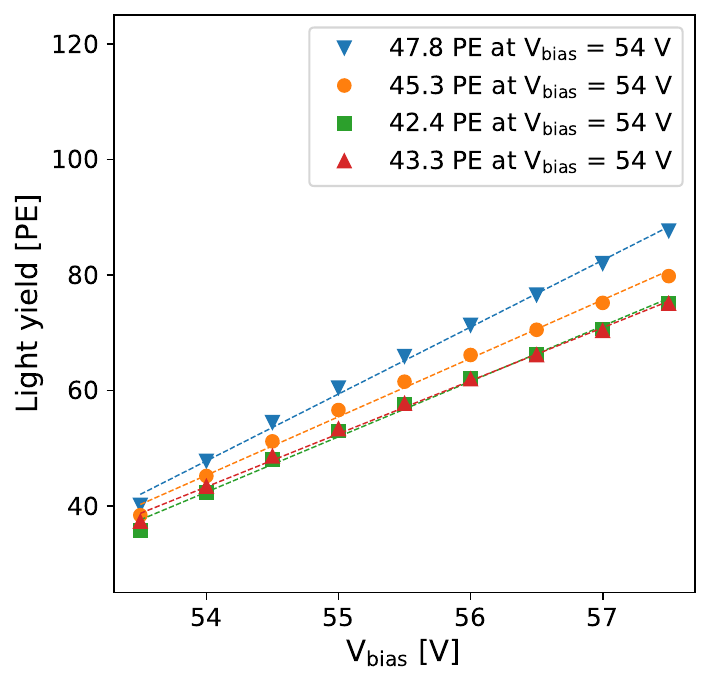}
    \label{subfig:PE_vs_bias}
  \end{subfigure}
  \begin{subfigure}[b]{0.49\textwidth}
    \includegraphics[width=0.9\linewidth]{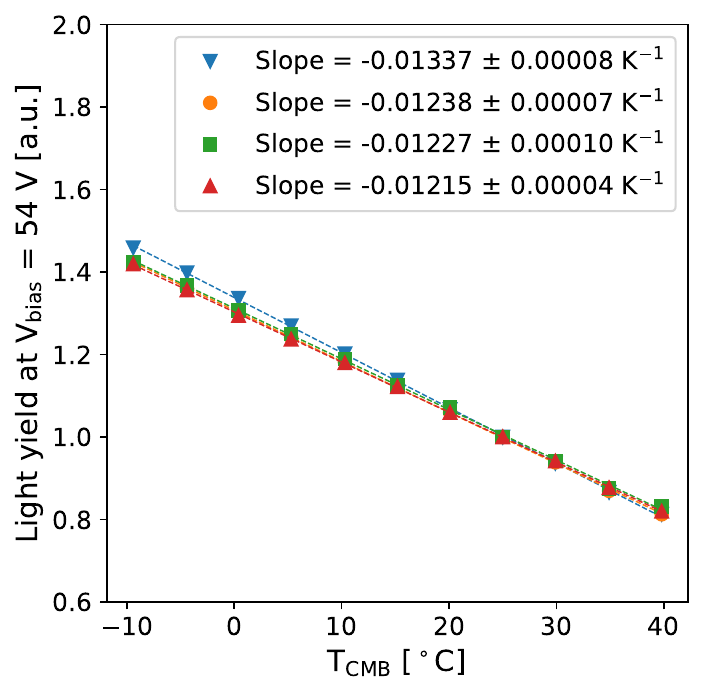}
    \label{subfig:PE_vs_temp_54V}
  \end{subfigure}
  \caption{Left:  SiPM response, in photoelectrons, to the flasher LED as a function of the bias voltage. The differences among the SiPMs are caused by the relative positions of the SiPMs relative to the LED. Right: Temperature dependence of the LED response of the SiPMs with the bias voltage set to 54.0\,V. The light yields have been normalized to be identical at 25$^{\circ}\mathrm{C}$.}
  \label{fig:PE_old_method}
\end{figure}

The SiPM response to the flasher LED light versus bias voltage, for runs taken under the same temperature, is shown in Fig.~\ref{fig:PE_old_method} left, resulting in a linear behavior. The SiPM response at a fixed bias voltage of 54.0\,V versus temperature is shown in Fig.~\ref{fig:PE_old_method} right.  That dependence is also linear, the average change in light yield being $(\text{-}1.27\pm$0.08)\%/K. Convoluting the temperature dependence of the breakdown voltage with the light yield dependence on the breakdown voltage gives the light yield dependence with respect to changes in the breakdown voltage: (21.8${\pm}$2.1)\%/V.  This is parameter $p_5$ in Table~\ref{tab:corrections}.

There are limitations to this method. Due to different normalization, the calculations are specific to the choice of bias voltage, which has been somewhat arbitrarily selected. At a fixed bias voltage, the SiPM's light yield dependence on the temperature is a manifestation of the over-voltage dependence of SiPM's photon detection efficiency and cross-talk. Due to the SiPM-to-SiPM difference in the breakdown voltage, the over-voltage and hence the normalization are not exactly uniform across all the SiPMs. In addition, the light emission of the LED used as the flasher also had a temperature dependence. Although the flasher was placed outside the temperature chamber, the external LED temperature varied slightly with the chamber temperature. This introduced a systematic uncertainty in the above analysis. The effect is relatively small, as proven in Sec.~\ref{sec:validation}.  Hence, the corrections using the parameters derived above work well. Readers wishing to make a more precise light yield correction can choose to produce a light yield versus over-voltage curve for a specific SiPM and apply the correction accordingly.

\subsection{FEB Temperature Corrections}
\begin{figure}
\centering
\includegraphics[width=0.8\textwidth]{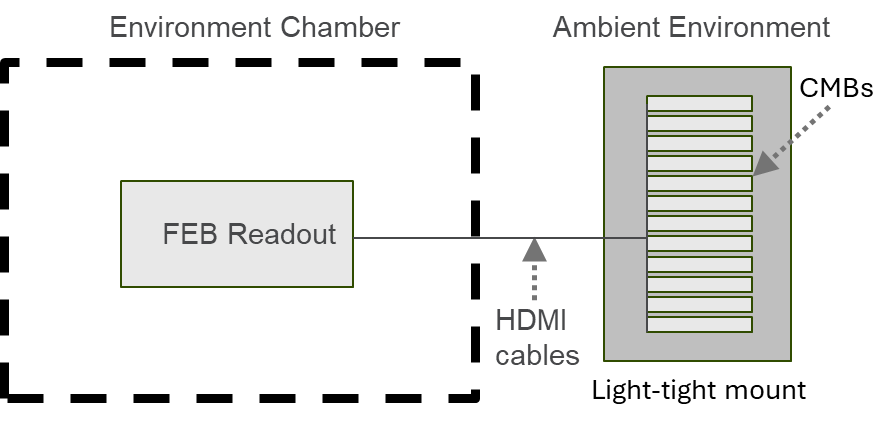}
\caption{Diagram of setup for FEB temperature correction studies at SiDet. The SiPMs were mounted, as before, on CMBs, four SiPMs for each CMB.  As shown in Fig.~\ref{fig:sidetpictures} there is a temperature sensor inside the chamber which can be monitored from the outside.}
\label{fig:febtempsetup}
\end{figure}

The bias provided to the SiPMs by the FEB, and the gain of the amplifier in the AFE chip of the FEB, are also temperature dependent, although these effects are about an order of magnitude smaller than those discussed above. Nevertheless, they remains non-negligible for precise measurements such as that required for aging studies.  To study these effects, we placed the readout FEB inside the environmental chamber and the SiPMs outside in a light-tight enclosure (Fig.~\ref{fig:febtempsetup}) . The SiPM biases were set to 55.0\,V. The voltages provided to the SiPMs and the dark pulses were measured for FEB board temperatures between 10$^{\circ}\mathrm{C}$ and 40$^{\circ}\mathrm{C}$. Note that the temperature sensor on the FEB typically measures a value about 7$^{\circ}\mathrm{C}$ above the set value of the environmental chamber due to heat from the components on the FEB board. 

\subsubsection{Bias Supply FEB Temperature Dependence}\label{sec:sipmvbdfebtemp}
\begin{figure}
\centering
\includegraphics[width=0.8\textwidth]{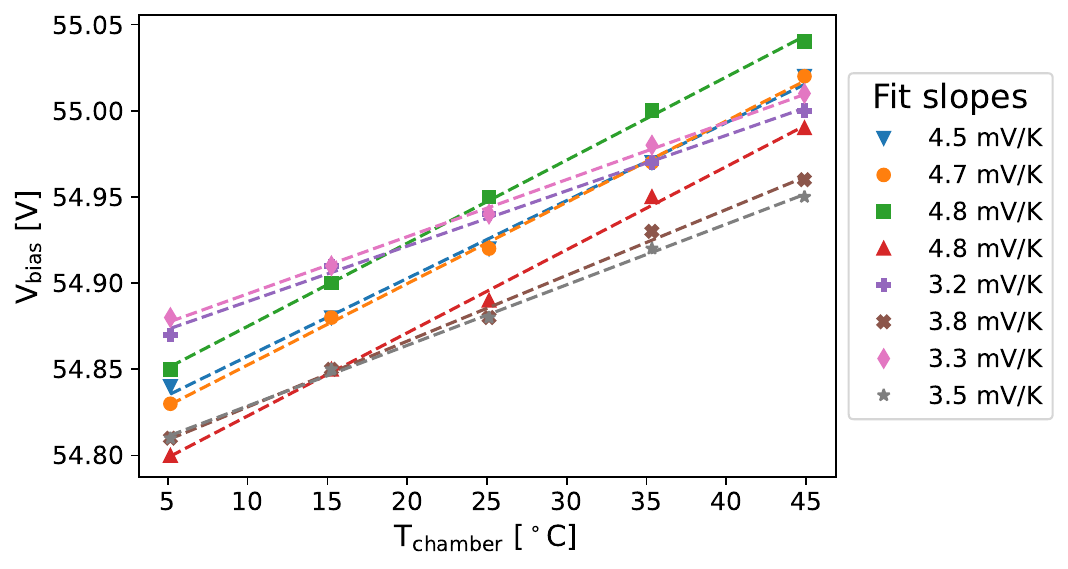}
\caption{SiPM bias voltages supplied by the FEB ($V_{\mathrm{bias}}$) versus environment chamber temperatures ($T_{\mathrm{chamber}}$), for 8 SiPMs. FEB bias settings were all at 55.0\,V. Note that the FEB temperatures, $T_{\mathrm{FEB}}$, are systematically higher than the chamber temperatures by about 7$^{\circ}\mathrm{C}$. The legend shows the slopes of the linear best-fit lines. The average is $dV_{\mathrm{bias}}/dT_{\mathrm{FEB}} = 4.1{\pm}0.7~\mathrm{mV/K}$.} \label{fig:FEB_bias_temp_impact}
\end{figure}

This correction accounts for temperature variations on the FEB board that affect the biases supplied to SiPMs.  This effect arises from the temperature sensitivity of the resistors used to provide the bias to the SiPMs.\footnote{In the upcoming FEB version, we plan to incorporate resistors with reduced temperature coefficients.} We measured the temperature coefficient of the FEB provided biases by placing the board in the environmental chamber and measuring at various chamber temperatures the voltages at the SiPMs using a hand-held multi-meter. On average the bias voltage increases by $4.1{\pm}0.7~\mathrm{mV/K}$ as the FEB temperature increases (Fig.~\ref{fig:FEB_bias_temp_impact}). This is parameter $p_2$ in Table~\ref{tab:corrections}.

\subsubsection{AFE Chip Gain FEB Temperature Dependence}
\begin{figure}
\centering
\includegraphics[width=0.48\textwidth]{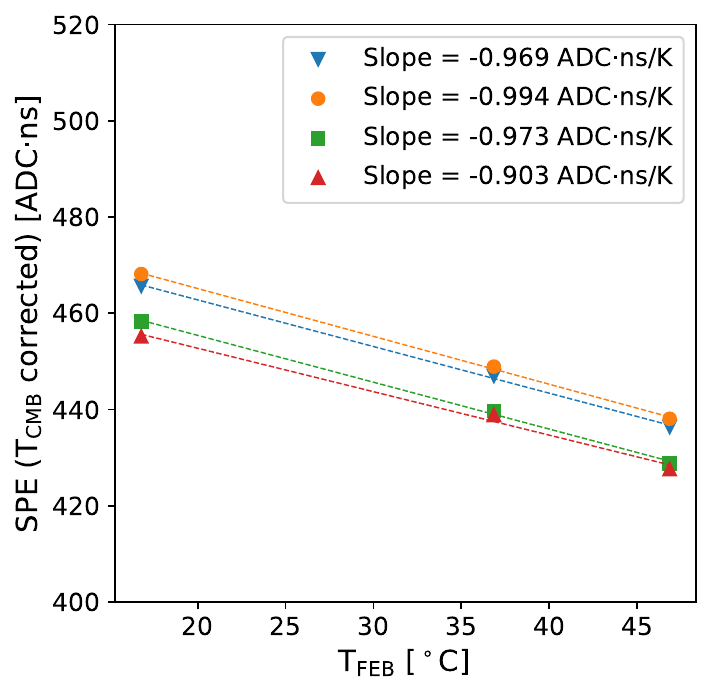}
\caption{SPE values versus the FEB temperature, after correcting for the effects of the small SiPM temperature variations.} \label{fig:SPE_corr_FEB_temp}
\end{figure}

Figure~\ref{fig:SPE_corr_FEB_temp} shows the SPE values of the four example SiPMs at different FEB temperatures, using the setup illustrated in Fig.~\ref{fig:febtempsetup}. Note that there were minor SiPM temperature differences, which were corrected as discussed in Sec.~\ref{sec:sipmtemp}. On average, a slope of $d\,{\rm SPE}/dT_{\rm FEB} = \text{-}0.95{\pm}0.13~\mathrm{ADC\cdot ns/K}$ was observed. This change is from the temperature dependence of the SiPM bias provided by the FEB, as well as the temperature dependence of the AFE amplifier of the FEB.  To determine the effect of the bias change, and remove it, we take $dV_{\mathrm{\rm bias}}/dT_{\mathrm{FEB}} = 4.1{\pm}0.7~\mathrm{mV/K}$ determined above and multiply it by parameter $p_3$ of Table~\ref{tab:corrections}, $d\,{\rm SPE}/dV_{\rm bias} = 125\pm$3 ADC${\cdot}$ns/V, to get $d\,{\rm SPE}/dT_{\rm FEB} = 0.51{\pm}0.10~\mathrm{ADC\cdot ns/K}$. Subtracting this off gives $d\,{\rm SPE}/dT_{\rm AFE} = \text{-}1.46{\pm}0.16~\mathrm{ADC\cdot ns/K}$: the temperature dependence of the AFE amplifier, which is the only other gain-related temperature dependence. This is parameter $p_4$ from Table~\ref{tab:corrections}. Note that this value is consistent with the information from the AFE data sheet. However, it should also be noted this value is specific to the gain setting we use for the AFE chip; a different setting will lead to a different correction.

\subsection{Parameter Validation} \label{sec:tempcorrectionsanitychecks}

To verify the calibration technique described above, in this section we first perform a validation check using the same data from which the parameters were derived.  In the following section we apply the technique to cosmic-ray data taken with the CRV modules at Wideband. 

\begin{figure}
  \centering
  \begin{subfigure}[b]{0.49\textwidth}
    \includegraphics[width=\linewidth]{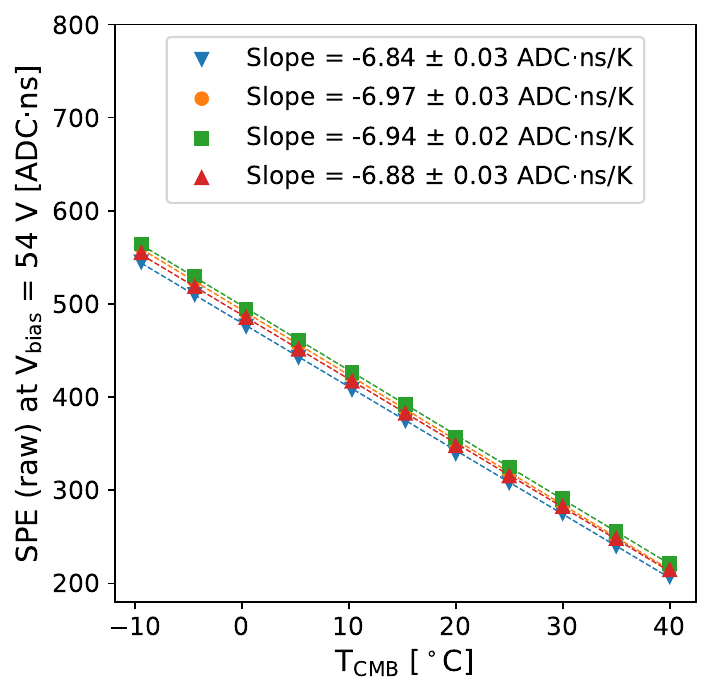}
  \end{subfigure}
  \begin{subfigure}[b]{0.49\textwidth}
    \includegraphics[width=\linewidth]{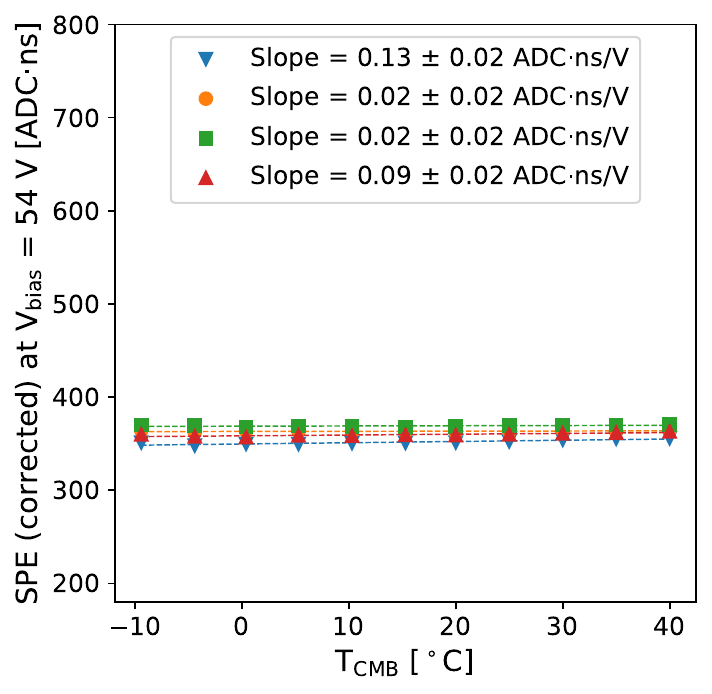}
  \end{subfigure}
  \caption{The SPE dependence on ambient temperature for the example SiPMs in our setup, before (left) and after (right) applying corrections.}
\label{fig:SPE_vs_temp}
\end{figure}

We reconstruct the SPE before and after the corrections, interpolating the SPE at a nominal bias of 54.0\,V (corresponding to an over-voltage of $\sim$2.5 V at $25^\circ$C) for different temperature runs. The before and after results are shown in Fig.~\ref{fig:SPE_vs_temp} for the four SiPMs. Before the temperature corrections, SPE temperature dependence was $-6.9{\pm}0.1 ~\mathrm{ADC \cdot ns/K}$, averaging over the four arbitrarily chosen SiPMs. After corrections, the temperature dependence was reduced to $0.2{\pm}0.1 ~\mathrm{ADC \cdot ns/K}$. 

\begin{figure}
\centering
\includegraphics[width=0.48\textwidth]{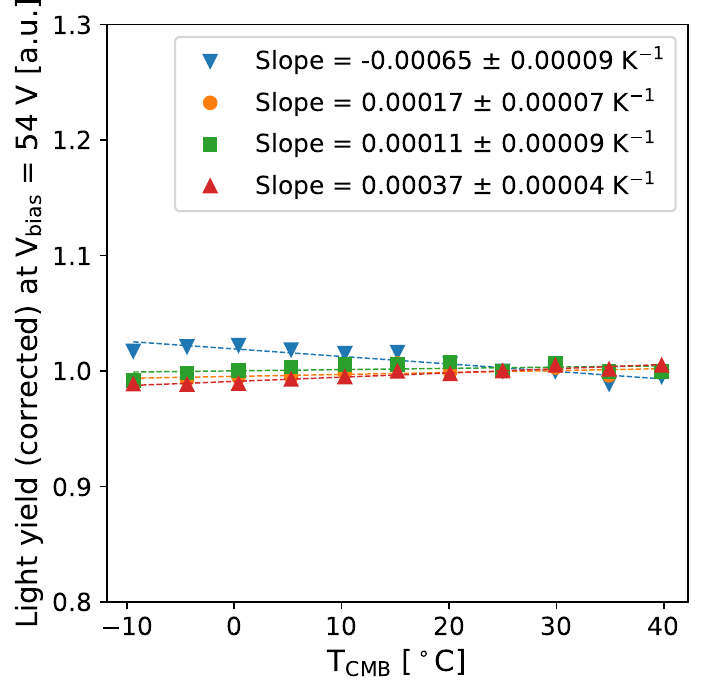}
\caption{Temperature-corrected LED light yield at 54.0\,V vs temperature after all temperature corrections have been applied for the example SiPMs. } 
\label{fig:PE_corr_vs_temp}
\end{figure}

We also measured the LED light yield at 54.0\,V bias voltage after the corrections, as shown in Fig.~\ref{fig:PE_corr_vs_temp}. This can be compared with the right plot in Fig.~\ref{fig:PE_old_method}, which shows the response of the same SiPMs before any corrections. When averaged over all channels, the residual slope is (0.1${\pm}$0.1)\%$/\mathrm{K}$, verifying that any remaining temperature dependence is small. The correction works well, and the LED light yields across a 50$^\circ$C temperature range vary by less than 5\%. Note that there are small SiPM-to-SiPM differences, and the deviations from perfectly flat lines in Fig.~\ref{fig:PE_corr_vs_temp} come from nonlinearity in the relation between SiPM light yield and over-voltage, as well as variations in the LED light emission: the determination of the parameter $p_5$ assumes the light from the LED was constant.\footnote{In our measurements, the LED temperature was controlled to better than a 10$^\circ$C range, corresponding to a light emission difference of less than 0.5\%.  To further improve this measurement, one should keep the LED and the SiPMs monitoring its light emission at a constant temperature.} Nonetheless, these deviations are comparably small. As shown in the next section, the correction scheme we derived here is sufficient for the typical data we collect since the temperature variations at the storage facility, and the expected variations during Mu2e operations, are much smaller.\footnote{Temperature measurements in the detector hall, without the top shielding installed and with doors to the detector hall occasionally open to the outside, have a maximum yearly variation of 6$^{\circ}$C. That variation is expected to be much less when the detector hall is closed for data taking. }

In principle, each calibration parameter in Table\,\ref{tab:corrections} can be measured individually for each SiPM and FEB channel, which would give even better temperature correction results. In practice, we chose to use the average values to perform corrections, because the impact from the SiPM and FEB channel-to-channel variations is small, and it is not feasible to measure all the parameters for the $\sim$20,000 SiPMs and FEB channels used in Mu2e. A large change of 10$^\circ$C in temperatures corresponds to less than a 1\% error in both the SPE calibration constant and the light yield determination. In addition, dark pulses will be taken concurrently with the signals under the same temperature conditions, and the impacts on the SPE calibration constant are largely canceled out when translating from $T^{\rm Dark}$ to $T^{\rm Meas}$. 

\section{Validation of Calibration Scheme with CRV Modules}
\label{sec:validation}

In this section we validate the method by applying it to different counters using cosmic-ray data collected with CRV modules stored at Wideband.

\begin{figure}
  \centering
  \includegraphics[width=0.75\linewidth]{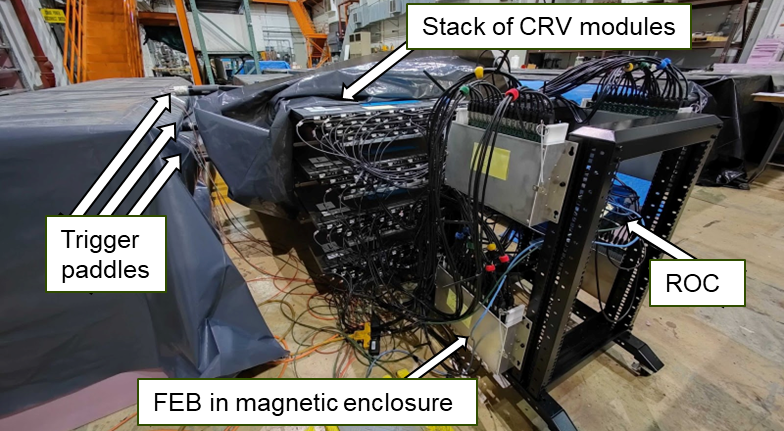}
    \label{subfig:Wideband_setup}
  \includegraphics[width=0.9\linewidth]{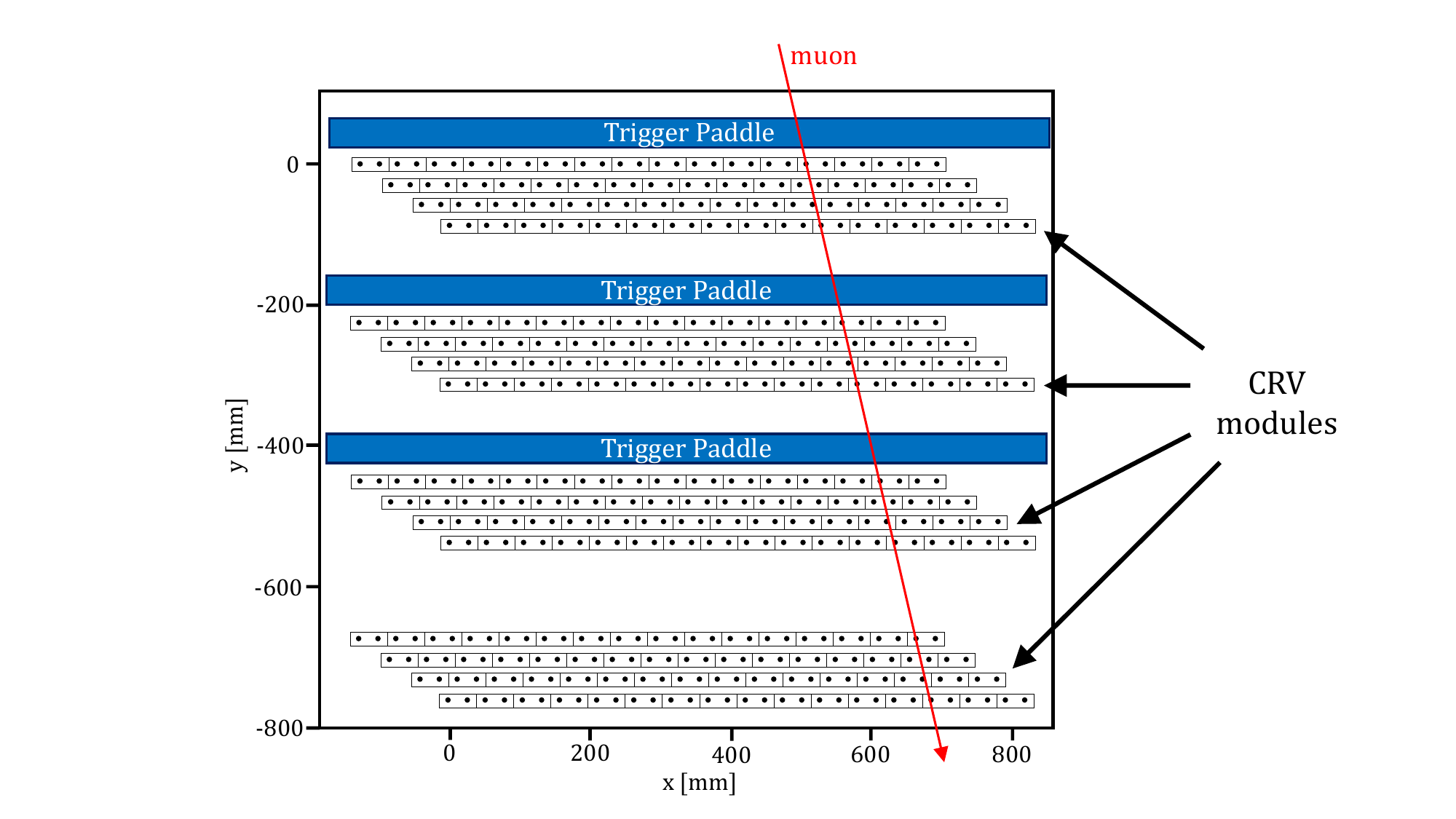}
    \label{subfig:Wideband_setup_cartoon}
    \caption{Top: Cosmic-ray test stand at the Wideband hall showing the trigger paddles and readout system (as described in Sect.~\ref{sec:crv} and~\ref{sec:readout}). Bottom: End view cartoon of the cosmic-ray test stand.}
  \label{fig:wideband}
\end{figure}

Modules at the Wideband detector hall are stored in stacks as shown in Fig.~\ref{fig:wideband}. Three plastic scintillator trigger paddles read out with photomultiplier tubes, were placed at one meter from the readout end, perpendicular to the length of the CRV counters. A 3-fold coincidence of signals in the trigger paddles within 100\,ns produced a trigger; primarily cosmic ray muons. The trigger was distributed to the FEBs through the ROC.

We analyzed three cosmic data runs, two experiencing similar ambient temperature conditions and the third conducted under significantly lower temperatures. These runs were examined to assess the impact of the temperature corrections described above.

\begin{figure}[!htb]
  \begin{subfigure}{0.49\textwidth}
    \includegraphics[width=\linewidth]{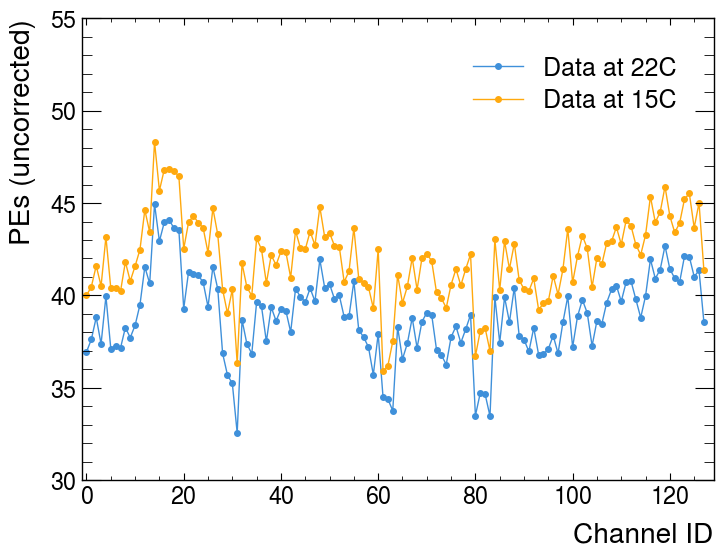}
  \end{subfigure}
  \begin{subfigure}{0.49\textwidth}
    \includegraphics[width=\linewidth]{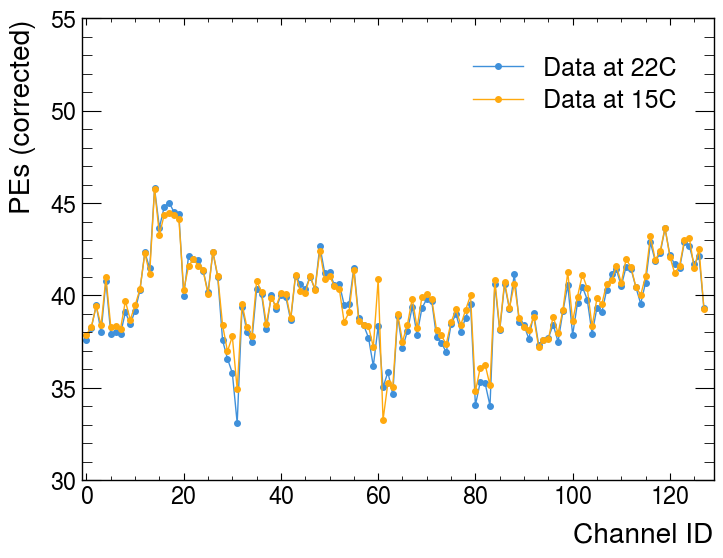}
  \end{subfigure}
  \caption{Left: Reconstructed SiPM light yields (in PEs) from cosmic-ray muons for two data runs, one taken at a room temperature $15^{\circ}$C (orange), the other at $22^{\circ}$C (blue), before correction.  Right: SiPM light yields, adjusted to the nominal SiPM temperature of $20^{\circ}$C. \label{fig:TempScan_PE}}
\end{figure}

The two runs taken with comparable ambient conditions, characterized by SiPM temperatures of approximately $22^\circ$C, resulted in light yield differences of $(0.3{\pm}2.0)$\%. The third run, performed at approximately $15^\circ$C, was compared to one of the $22^\circ$C runs.  The uncorrected and temperature-corrected photoelectron yields are shown in Fig.~\ref{fig:TempScan_PE}. With a temperature difference of $7^\circ$C between the two runs, our correction algorithms resulted in a light yield difference of $(-0.5{\pm}1.3)$\% when averaged over all 64 SiPMs. 

\begin{figure}[ht]
\centering
\includegraphics[width=0.6\linewidth]{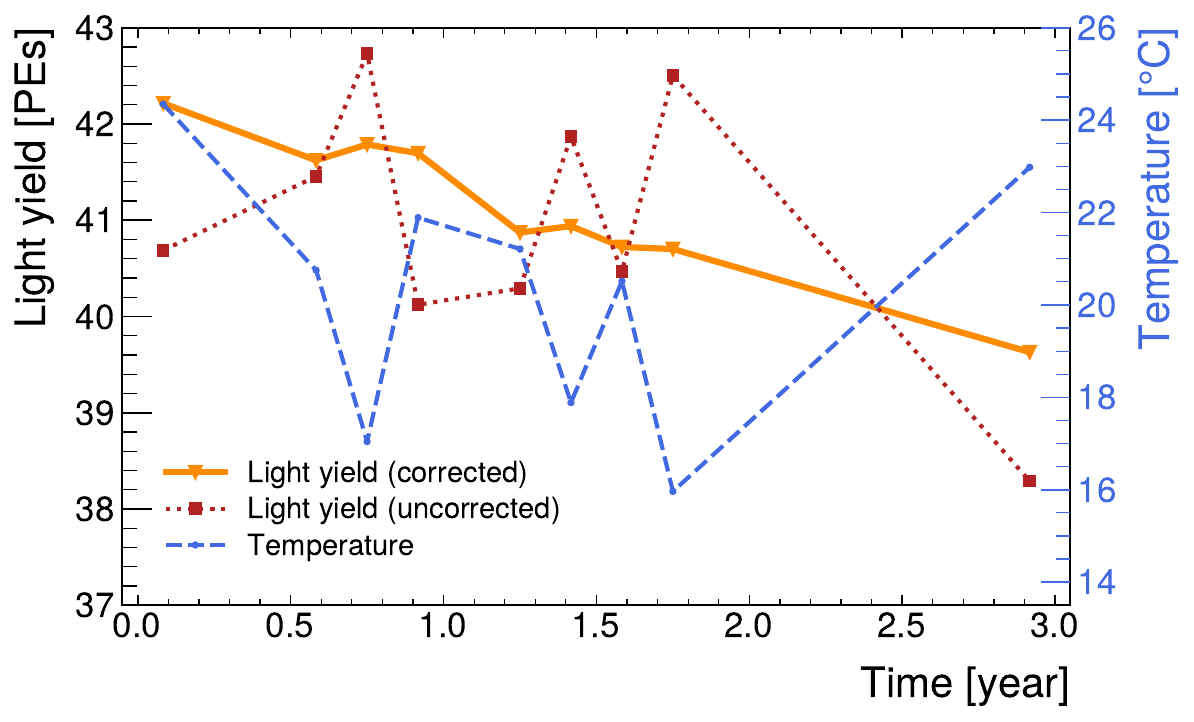}
\caption{The light-yield response of a sample SiPM to cosmic-ray muons before and after the temperature corrections,  measured at nine times over a three-year period at different temperatures.  Without the corrections it would be difficult to measure the aging of the CRV counters to the required precision.  \label{fig:AgingExample}}
\end{figure}

As a final validation of this method, we show the light yield response to cosmic-ray muons for an example CRV SiPM over three years in Fig.~\ref{fig:AgingExample}.  Without temperature corrections, the light yield varies inversely with the temperature.  After the correction, the variations are mitigated and there is a clear decline in the light yield with time, or aging of the scintillator.  A complete analysis of CRV aging results will be presented in a forthcoming presentation.  

Currently, the temperature at the Mu2e hall has a maximum yearly variation of $6^\circ$C. When the detector hall is closed up for data taking it is expected to have less variation during Mu2e operations with the shielding in place and the doors to the detector hall closed. The stable operation environment combined with the temperature correction scheme described above will ensure a consistent performance of the CRV system.

\section{Summary}\label{sec:summary}
A method for calibration and handling the temperature dependence of non-temperature controlled SiPMs used for the Mu2e cosmic ray veto detector (CRV) has been described in detail.  The method corrects the single photoelectron measurement to a reference temperature to allow comparison of measurements taken at different temperatures.  
For this to work, the temperature dependence of all of the components affecting the readout of the SiPMs must be well understood:  hence, the temperature dependence measurements were made in an environmental chamber.  The method was validated using an independent experimental setup with cosmic-ray muon data collected using CRV modules.  When the results from data samples were taken from two runs with large temperature differences ($\Delta T=7^{\circ}$C) the corrected results were found to be consistent, that is no residual dependence on temperature was observed.

\section{Acknowledgements}\label{sec:acknowledgements}
We are grateful for the vital contributions of the Fermilab staff and the technical staff of the participating institutions. This work was supported by the US Department of Energy; the Istituto Nazionale di Fisica Nucleare, Italy; the Science and Technology Facilities Council, UK; the Ministry of Education and Science, Russian Federation; the National Science Foundation, USA; the National Science Foundation, China; the Helmholtz Association, Germany; and the EU Horizon 2020 Research and Innovation Program under the Marie Sklodowska-Curie Grant Agreement Nos.  734303, 822185, 858199, 101003460, and 101006726. This document was prepared by members of the Mu2e Collaboration using the resources of the Fermi National Accelerator Laboratory (Fermilab), a U.S. Department of Energy, Office of Science, HEP User Facility. Fermilab is managed by Fermi Research Alliance, LLC (FRA), acting under Contract No. DE-AC02-07CH11359.

\bibliographystyle{unsrt}  

\bibliography{references} 

\newpage
\pagenumbering{arabic}

\end{document}